\documentclass[
aps,prb,twocolumn,amsmath,amssymb,floatfix,
superscriptaddress,longbibliography,footinbib
]{revtex4-2}

\usepackage[dvips]{graphics}
\usepackage{bm}
\usepackage{float}
\usepackage{epsfig}
\usepackage{enumerate}
\usepackage{amsmath}
\usepackage{color}
\usepackage{graphicx}
\usepackage[colorlinks=true,linktoc=page,linkcolor=red,citecolor=blue]{hyperref}
\usepackage{natbib}
\usepackage{float}
\usepackage{graphicx}
\usepackage{dcolumn}
\usepackage{bm}
\usepackage{latexsym,epsfig}
\usepackage{graphicx}
\usepackage{verbatim}
\usepackage{comment}
\usepackage{amsmath}
\usepackage[utf8]{inputenc}
\usepackage{amssymb}
\usepackage{physics}
\usepackage{stmaryrd}
\usepackage{color}
\usepackage{epstopdf}
\usepackage{grffile}
\usepackage{lipsum}
\usepackage{enumitem}
\usepackage[toc,page]{appendix}
\usepackage[dvipsnames]{xcolor}
\DeclareGraphicsExtensions{.eps}

\newcommand{\beq}{\begin{equation}}
\newcommand{\eeq}{\end{equation}}
\newcommand{\bea}{\begin{eqnarray}}
\newcommand{\eea}{\end{eqnarray}}
\newcommand{\ben}{\begin{eqnarray*}}
\newcommand{\een}{\end{eqnarray*}}
\newcommand{\bfig}{\begin{figure}}
\newcommand{\efig}{\end{figure}}
\usepackage{soul}

\usepackage{hyperref}
\hypersetup{
    colorlinks=true,      
    urlcolor=blue,
    citecolor=blue,
    linkcolor=blue
}

\begin{document}

\title{Multiple Current Enhancement Regimes and Interaction Induced Re-entrant Transport in a Quasiperiodic Hubbard Ring with Antiferromagnetic Order}






\author{Souvik Roy}
\email{Corresponding author: souvikphysics5293@gmail.com}
\affiliation{School of Basic Sciences, Indian Institute of Technology Bhubaneswar, Odisha 752050, India}

\author{Ranjini Bhattacharya}
\email{ranjinibhattacharya@gmail.com}
\affiliation{Institute of Physics, Sachivalaya Marg, Bhubaneswar-751005, India}
\affiliation{Homi Bhabha National Institute, Training School Complex, Anushaktinagar, Mumbai 400094, India}
\email{souvikphysics5293@gmail.com}

\date{\today}

\begin{abstract}
We investigate persistent charge and spin currents in a one-dimensional quantum ring threaded by an Aharonov-Bohm (AB) flux, considering generalized Aubry-Andr\'e-Harper (AAH) modulation with staggered onsite energies and antiferromagnetic (AFM) ordering in both the noninteracting and interacting regimes, with the latter treated using a self-consistent Hartree-Fock mean field theory (MFT). At half filling, two distinct current-enhancement peaks emerge as the modulation strength increases, with quasiperiodic hopping modulation playing a crucial role in establishing the two-peak structure within a restricted parameter regime. Spin-resolved bond-current profiles reveal a switch in the dominant spin contribution between the two peaks, while the inverse participation ratio (IPR) and flux-dependent spin-resolved ground-state energies provide complementary insights into the spatial characteristics of the eigenstates and their spin-dependent flux sensitivity. At quarter filling, the spin current exhibits a single enhancement peak, in contrast to the two-peak response observed at half filling. The Hubbard interaction affects the two half-filled peaks differently: it suppresses the primary peak, whereas the secondary peak exhibits a nonmonotonic, re-entrant response, with the current recovering at finite interaction strength. The interacting results are further characterized through the Drude weight, and comparisons with exact diagonalization (ED) and density-matrix renormalization group (DMRG) calculations for a finite half-filled system support the reliability of the MFT description. Our findings elucidate how quasiperiodic modulation, AFM order, band filling, and electronic correlations collectively govern the magnitude and spin-resolved character of persistent currents, giving rise to multiple enhancement regimes and interaction-induced re-entrant behavior.
\end{abstract}

\maketitle
\section{Introduction}
\label{sec:introduction}

Quantum transport in low-dimensional systems is strongly influenced by the interplay between electronic coherence, spatial inhomogeneity, and electron-electron interactions. In one dimension, disorder can profoundly modify the spatial structure of electronic eigenstates~\cite{r1,r3,r2}, often suppressing their ability to support extended transport~\cite{r6,r5,r4}. Deterministic quasiperiodic potentials, however, provide an intermediate setting between crystalline order and random disorder, in which competing length scales generate intricate spectral structures and nontrivial localization properties. The Aubry-Andr\'e-Harper (AAH)~\cite{r8,r7,r9,r12,r11,r10, r24, r26, r28, r30,r32, r34, r36, r38} family of models has emerged as a central framework for investigating these effects, particularly because the modulation strength can drive substantial changes in the spatial character of the eigenstates~\cite{r12,r14, r17,r15,r16,r19,r17,r20,r18,r25,r27,r29,r31,r33,r35,r37,r39}. When such modulation is introduced into a finite quantum ring threaded by an Aharonov-Bohm (AB)~\cite{r40,r41,r42} flux, the resulting system provides a natural platform for exploring how quasiperiodicity influences coherent circulation and the magnetic-flux sensitivity of electronic states.


Persistent currents~\cite{r43,r46,r45,r44} provide a useful probe of the interplay between quasiperiodicity, hopping modulation, and electronic correlations in mesoscopic rings. The magnetic-flux dependence of the ground-state energy determines the persistent current~\cite{r48,r49,r51,r47,r50}, thereby capturing both the energy spectrum and the flux sensitivity of the occupied states~\cite{r54,r52,r53}. In a ring with only an AAH site potential, increasing the quasiperiodic modulation generally suppresses the persistent current~\cite{dey,nilanjan}. By contrast, an appropriate staggered-non-staggered arrangement of site energies combined with hopping dimerization can enhance the current over a finite parameter range, although further modulation eventually reduces it~\cite{souvik3}. Similarly, incorporating a Fibonacci sequence at the unit-cell level alongside hopping dimerization can yield an enhancement of the persistent current with increasing Fibonacci modulation strength~\cite{souvik1}. These findings demonstrate that current suppression or enhancement depends sensitively on the spatial organization of the site energies and hopping amplitudes. However, the possibility of multiple distinct current-enhancement regimes within a single modulation scan appears to remain comparatively unexplored.

The interplay of Hubbard interactions~\cite{r56,r58,r57,r59,r60,r61,r62,r65,r64,r63}, quasiperiodic modulation, and antiferromagnetic ordering provides a rich setting for investigating persistent charge and spin currents in interacting quantum rings. Previous studies have established that electron-electron interactions can enhance persistent currents under suitable conditions~\cite{souvikmft}, demonstrating that correlations need not lead solely to current suppression. However, the possibility of a re-entrant current response, in which an initial suppression is followed by a recovery at finite interaction strength, remains comparatively unexplored in systems combining quasiperiodicity and staggered magnetic ordering. Moreover, the Hubbard interaction couples the two spin sectors through their local occupations, while antiferromagnetic ordering introduces a staggered, spin-dependent energy landscape. Their combined influence can therefore produce distinct responses in the charge and spin currents. This motivates a systematic investigation of nonmonotonic persistent-current responses and the underlying spectral rearrangements, with particular emphasis on how quasiperiodic modulation, hopping structure, magnetic ordering, and Hubbard interactions jointly govern coherent charge and spin circulation in low-dimensional quantum systems.

In this work, we explore how quasiperiodic modulations, magnetic ordering, and electron-electron interactions influence persistent charge and spin currents in an AB ring of spinful fermions. The model combines staggered Aubry-André-Harper (AAH) site potentials with quasiperiodically modulated hopping amplitudes, antiferromagnetic (AFM) order, and an on-site Hubbard interaction. To investigate the interacting regime, we employ a self-consistent Hartree–Fock mean-field theory (MFT). Particular attention is paid to the dependence of the current response on particle filling and hopping-modulation strength, as well as to the interplay between the two spin sectors. We use spin-resolved bond currents to examine the microscopic origin of the current response, while the inverse participation ratio (IPR) and flux-dependent ground-state energies offer complementary perspectives on the eigenstates and their magnetic-flux dependence. The analysis also considers the Drude weight to characterize the response of the interacting system and benchmarks the mean-field results against ED and DMRG calculations for a finite half-filled ring. This framework allows us to examine the distinct roles of quasiperiodic site and hopping modulations, AFM order, and particle filling in shaping the current response. We further investigate how increasing the Hubbard interaction modifies the competing current-enhancement features, particularly the suppression and subsequent recovery of the secondary peak. These considerations provide a basis for understanding the emergence of multiple current-enhancement regimes and re-entrant behavior in correlated quasiperiodic quantum rings.

The remainder of this paper is organized as follows. Section~II presents the model Hamiltonian and the theoretical framework. Section~III is divided into two parts, addressing the noninteracting and interacting regimes separately. Section~IV summarizes the main findings and discusses possible directions for future research. Appendix~A provides a complementary analysis of the ground-state energy variations.

\section{Model and Theoretical Formulation}
In this section, we present the model Hamiltonian and theoretical framework for the quantum ring geometry considered in this work. Beginning with the Hamiltonian formulation, we develop the expressions for the relevant transport properties, including the persistent charge (spin) currents and the Drude weight. We also describe the self-consistent procedure employed to incorporate the Hubbard interaction within the Hartree-Fock mean-field approximation.
\label{sec:model}
\begin{figure}[t]
    \centering
    \includegraphics[width=8cm,height=5cm]{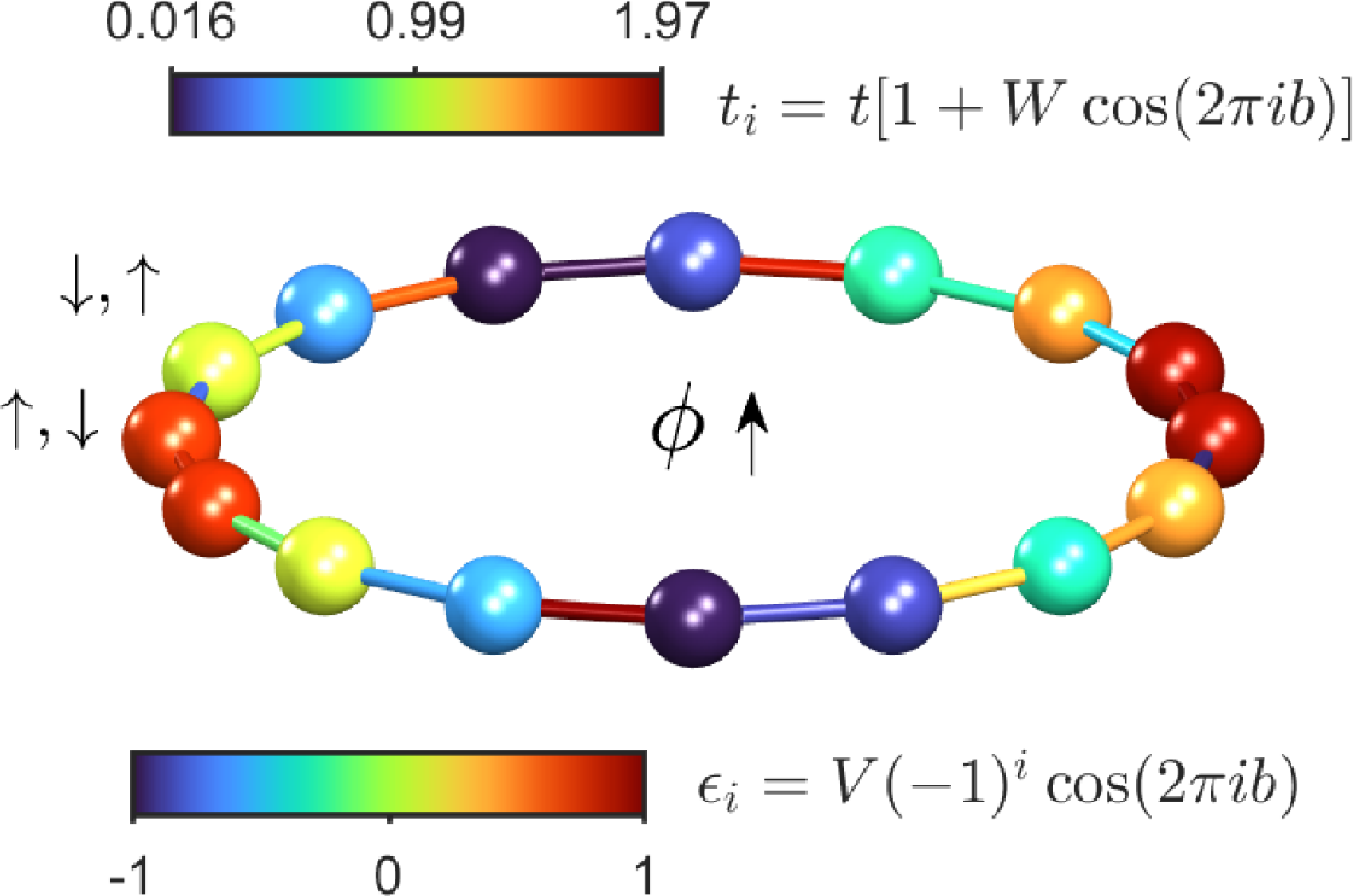}
    \caption{Schematic representation of a one-dimensional quasiperiodic quantum ring with staggered AAH site-energy modulation and quasiperiodically modulated hopping amplitudes. The site colors represent the staggered AAH potential, $\epsilon_i=V(-1)^i\cos(2\pi i b)$, while the bond colors indicate the hopping modulation, $t_i=t[1+W\cos(2\pi i b)]$, with $b=(\sqrt{5}-1)/2$. The arrows at sites 1 and 2 indicate the spin configurations $(\uparrow,\downarrow)$ and $(\downarrow,\uparrow)$, respectively. The two color scales distinguish the site-energy and hopping modulations}
    \label{fig:pic}
\end{figure}
\subsection{Hamiltonian}

We consider a one-dimensional ring (Fig.~\ref{fig:pic}) of $N$ sites occupied by spin-$1/2$
fermions. The ring is threaded by an Aharonov-Bohm (AB) flux, while
the hopping amplitudes and on-site energies are spatially modulated.
In addition, a staggered Zeeman field and an on-site Hubbard interaction
are included to examine the combined effects of quasiperiodicity,
magnetic ordering, and electronic correlations on persistent charge
and spin currents. The Hamiltonian is written as
\begin{equation}
\begin{aligned}
H ={}&
-\sum_{i=1}^{N}\sum_{\sigma}
t_i\left(
e^{i\theta}c_{i\sigma}^{\dagger}c_{i+1,\sigma}
+\mathrm{H.c.}
\right)
\\
&+\sum_{i=1}^{N}\sum_{\sigma}
\left[
V_i+\sigma h_z(-1)^{i}
\right]n_{i\sigma}
\\
&+U\sum_{i=1}^{N}n_{i\uparrow}n_{i\downarrow},
\end{aligned}
\label{eq:hamiltonian}
\end{equation}
where $c_{i\sigma}^{\dagger}$ and $c_{i\sigma}$ are the creation and
annihilation operators for an electron with spin
$\sigma\in\{\uparrow,\downarrow\}$ at site $i$, and
$n_{i\sigma}=c_{i\sigma}^{\dagger}c_{i\sigma}$ is the corresponding
number operator. Periodic boundary conditions are imposed,
$c_{N+1,\sigma}=c_{1,\sigma}$, so that the final bond closes the ring.
We use $\sigma=+1$ for spin-up electrons and $\sigma=-1$ for
spin-down electrons.

We consider a generalized AAH model~\cite{r27} in which both the nearest-neighbor hopping amplitudes and the on-site energies are modulated quasiperiodically. The hopping amplitude is given by
\begin{equation}
t_i=t\left[1+W\cos\left(2\pi\beta i \right)\right],
\label{eq:hopping}
\end{equation}
where $t$ sets the characteristic hopping-energy scale, $W$ controls the hopping-modulation strength, and $\beta=(\sqrt{5}-1)/2$ is the inverse golden ratio. The on-site potential is described by a staggered AAH modulation,
\begin{equation}
V_i=V\cos(2\pi\beta i)(-1)^{i},
\label{eq:onsite}
\end{equation}
where $V$ denotes the potential-modulation amplitude. Here, the cosine term introduces an incommensurate spatial modulation, while the factor $(-1)^{i}$ imposes an alternating sign between neighboring sites. This combination generates a quasiperiodic potential with an additional staggered structure, distinguishing it from the conventional AAH potential without the alternating factor~\cite{r25,r27,r26,r31,r35}. The interplay between the modulated hopping and the staggered on-site potential provides a means of tuning the electronic energy landscape, potentially influencing the spatial localization of the eigenstates and the persistent-current response.

In addition, a staggered Zeeman field of amplitude $h_z$ acts with opposite signs on the two spin species and alternates between neighboring sites. Its contribution to the local energy is therefore spin dependent, favoring opposite spin orientations on the two sublattices. The combined effect of the staggered AAH potential and the staggered Zeeman field produces a spatially and spin-dependent on-site energy landscape. Their interplay can modify the spin-resolved eigenstates and lead to distinct responses in the charge and spin persistent currents.

The AB flux is incorporated through the Peierls phase
\begin{equation}
\theta=\frac{2\pi\phi}{N},
\label{eq:peierls}
\end{equation}
where $\phi$ is the flux measured in units of the flux quantum
$\phi_0=h/e$. This convention distributes the total phase
$2\pi\phi$ uniformly over the $N$ bonds of the ring. The final term
in Eq.~\eqref{eq:hamiltonian} describes the on-site Hubbard interaction
between electrons of opposite spin, with $U$ denoting the interaction
strength.

\subsection{Self-consistent mean-field formulation}

The Hubbard term in Eq.~\eqref{eq:hamiltonian} couples the two spin
species through their local occupations. To investigate the system
over the sizes considered here, we employ a self-consistent
Hartree-Fock mean-field approximation~\cite{r56}. Within this treatment, the
interaction term is decoupled as
\begin{equation}
\begin{aligned}
n_{i\uparrow}n_{i\downarrow}\simeq{}&
\langle n_{i\uparrow}\rangle n_{i\downarrow}
+\langle n_{i\downarrow}\rangle n_{i\uparrow}
\\
&-\langle n_{i\uparrow}\rangle
 \langle n_{i\downarrow}\rangle.
\end{aligned}
\label{eq:mf_decoupling}
\end{equation}
Here, the local densities are determined self-consistently from the
occupied single-particle states. The resulting effective Hamiltonian
can be expressed as
\begin{equation}
H_{\mathrm{MF}}=
H_{\uparrow}^{\mathrm{MF}}+
H_{\downarrow}^{\mathrm{MF}}-
U\sum_i
\langle n_{i\uparrow}\rangle
\langle n_{i\downarrow}\rangle,
\label{eq:mf_hamiltonian}
\end{equation}
where the spin-resolved Hamiltonians are
\begin{equation}
\begin{aligned}
H_{\uparrow}^{\mathrm{MF}}
={}&-\sum_i t_i
\left(e^{i\theta}c_{i\uparrow}^{\dagger}c_{i+1,\uparrow}
+\mathrm{H.c.}\right)
\\
&+\sum_i
\left[
V_i+h_z(-1)^{i}
+U\langle n_{i\downarrow}\rangle
\right]n_{i\uparrow},
\end{aligned}
\label{eq:mf_up}
\end{equation}
and
\begin{equation}
\begin{aligned}
H_{\downarrow}^{\mathrm{MF}}
={}&-\sum_i t_i
\left(e^{i\theta}c_{i\downarrow}^{\dagger}c_{i+1,\downarrow}
+\mathrm{H.c.}\right)
\\
&+\sum_i
\left[
V_i-h_z(-1)^{i}
+U\langle n_{i\uparrow}\rangle
\right]n_{i\downarrow}.
\end{aligned}
\label{eq:mf_down}
\end{equation}
The interaction therefore enters each spin sector through an effective
potential determined by the local density of the opposite spin.
The last term in Eq.~\eqref{eq:mf_hamiltonian} corrects for the
double counting of the interaction contribution introduced by the
mean-field decoupling.

The self-consistent solution is obtained iteratively. Initial
distributions for $\langle n_{i\uparrow}\rangle$ and
$\langle n_{i\downarrow}\rangle$ are used to construct the two
effective Hamiltonians, which are then diagonalized independently.
The updated densities are calculated from the occupied eigenstates,
\begin{equation}
\langle n_{i\sigma}\rangle
=\sum_{k=1}^{N_\sigma}
f_{k\sigma}|\psi_{k\sigma}(i)|^2,
\label{eq:density}
\end{equation}
where $\psi_{k\sigma}(i)$ is the amplitude of eigenstate $k$ at site
$i$, $N_\sigma$ is the number of occupied states in spin sector
$\sigma$, and $f_{k\sigma}$ is the occupation number. At zero
temperature, $f_{k\sigma}$ equals unity for an occupied state and
zero otherwise. The density-update steps are repeated until the
change in the local densities falls below a prescribed convergence
tolerance. For the calculations reported here, the total electron number is
fixed by the chosen filling. In the implementation, the same number
of states is occupied in each spin sector. The converged densities
define the effective Hamiltonians used to evaluate the energy,
persistent currents, and bond-resolved currents.

Having obtained the converged spin densities and the corresponding self-consistent Hamiltonian, we can now evaluate the relevant transport properties, as detailed in the following subsections.

\subsection{Ground-state energy and persistent currents}

At zero temperature, the mean-field ground-state energy is evaluated
from the occupied single-particle eigenvalues, with the interaction
double-counting correction included:
\begin{equation}
\begin{aligned}
E_0={}&
\sum_{k=1}^{N_\uparrow}\varepsilon_{k\uparrow}
+\sum_{k=1}^{N_\downarrow}\varepsilon_{k\downarrow}
\\
&-U\sum_i
\langle n_{i\uparrow}\rangle
\langle n_{i\downarrow}\rangle,
\end{aligned}
\label{eq:ground_energy}
\end{equation}
where $\varepsilon_{k\sigma}$ denotes the $k$th eigenvalue of the
converged spin-dependent mean-field Hamiltonian.

The persistent current is related to the sensitivity of the
ground-state energy to the applied AB flux. For the convention used
in Eq.~\eqref{eq:peierls}, the charge current is~\cite{souvik1}
\begin{equation}
I_c=-\frac{\partial E_0}{\partial\phi}.
\label{eq:charge_current}
\end{equation}
The spin-resolved contributions are obtained from the flux derivatives
of the occupied eigenvalues,
\begin{equation}
I_\sigma=-\sum_{k=1}^{N_\sigma}
\frac{\partial\varepsilon_{k\sigma}}{\partial\phi},
\label{eq:spin_current_components}
\end{equation}
where the derivatives are evaluated at the converged self-consistent
solution. With the convention adopted here, the spin current is
defined as the difference between the two spin-channel currents,
\begin{equation}
I_s=I_\uparrow-I_\downarrow,
\label{eq:spin_current}
\end{equation}
whereas the charge current is
\begin{equation}
I_c=I_\uparrow+I_\downarrow.
\label{eq:charge_current_spin}
\end{equation}
Overall prefactors associated with the choice of flux units and
physical current units are understood to be included when converting
the numerical results to current units.

\subsection{Bond-resolved current}

The total persistent current characterizes the net circulation around
the ring, but it does not by itself reveal how the current is
distributed among individual bonds. To resolve this spatial
structure, we calculate the current on each nearest-neighbor bond
$(i,i+1)$ separately for the two spin species. This quantity provides
a direct means of comparing the local contributions to circulation
at different modulation strengths.

For a normalized single-particle eigenstate $\psi_{k\sigma}$, the
bond-current~\cite{sudin} contribution is defined, in the convention used for the
numerical evaluation, by
\begin{equation}
J_{i,i+1}^{(k,\sigma)}
=2\,\operatorname{Im}\left[
\psi_{k\sigma}^{*}(i)
H_{i,i+1}^{\sigma}
\psi_{k\sigma}(i+1)
\right],
\label{eq:bond_current_state}
\end{equation}
where $H_{i,i+1}^{\sigma}$ is the nearest-neighbor matrix element of
the converged spin-dependent Hamiltonian. For the hopping term in
Eq.~\eqref{eq:hamiltonian},
\begin{equation}
H_{i,i+1}^{\sigma}=-t_i e^{i\theta}.
\label{eq:bond_hopping_matrix}
\end{equation}
The ring boundary is treated periodically, such that site $N+1$
corresponds to site $1$. The bond current in each spin sector is
obtained by summing over all occupied states:
\begin{equation}
J_{i,i+1}^{\sigma}
=\sum_{k=1}^{N_\sigma}
J_{i,i+1}^{(k,\sigma)}.
\label{eq:bond_current_total}
\end{equation}
Thus, $J_{i,i+1}^{\uparrow}$ and $J_{i,i+1}^{\downarrow}$ describe
the spin-resolved current distributions around the ring. Their sum
and difference define the bond-resolved charge and spin currents,
respectively:
\begin{align}
J_{i,i+1}^{c}
&=J_{i,i+1}^{\uparrow}
 +J_{i,i+1}^{\downarrow},
\label{eq:bond_charge_current}\\
J_{i,i+1}^{s}
&=J_{i,i+1}^{\uparrow}
 -J_{i,i+1}^{\downarrow}.
\label{eq:bond_spin_current}
\end{align}
The bond-current profiles reveal the relative contributions of the two spin channels. We use them alongside the flux derivative of the ground-state energy to characterize the persistent current and its spatial distribution.

\subsection{Drude Weight}
\label{subsec:drude_weight}

At zero temperature, the Drude weight characterizes the charge stiffness of the system and quantifies its ability to sustain dissipationless transport. In the absence of an applied magnetic flux, $\phi=0$, it is determined by the curvature of the ground-state energy with respect to the magnetic flux and is given by~\cite{souvik1,skm}
\begin{equation}
D_W=\left.\frac{N}{4\pi^2}
\frac{\partial^2 E_0}{\partial\phi^2}\right|_{\phi=0},
\label{eq:drude_weight}
\end{equation}
where $N$ denotes the number of lattice sites and $E_0$ represents the ground-state energy. This relation provides a measure of the sensitivity of the ground-state energy to an infinitesimal flux and offers a useful diagnostic for examining the influence of quasiperiodic modulation and Hubbard interactions on charge transport.

\subsection{Spin-Resolved Inverse Participation Ratio}
\label{subsec:IPR}

To characterize the spatial localization properties of the spin-resolved eigenstates, we calculate the inverse participation ratio (IPR) separately for the spin-up and spin-down sectors. For a normalized eigenstate $|\psi_{n,\sigma}\rangle=\sum_{i=1}^{N}\psi_{n,\sigma}(i)|i\rangle$, the spin-resolved IPR is defined as~\cite{r25,soumya1,soumya2}
\begin{equation}
\mathrm{IPR}_{n,\sigma}=\sum_{i=1}^{N}|\psi_{n,\sigma}(i)|^4,
\label{eq:IPR}
\end{equation}
where $n$ labels the eigenstate, $\sigma=\uparrow,\downarrow$ denotes the spin projection, and $\psi_{n,\sigma}(i)$ is the corresponding wave-function amplitude at site $i$. The eigenstates satisfy the normalization condition $\sum_{i=1}^{N}|\psi_{n,\sigma}(i)|^2=1$. For an extended state, the IPR scales as $1/N$, whereas for a localized state, it remains finite in the thermodynamic limit. Thus, smaller IPR values indicate greater spatial extension, while larger values signify stronger localization. The spin-resolved IPR provides a complementary measure for investigating how quasiperiodic modulation and staggered magnetic ordering influence the localization properties of the two spin sectors.

\maketitle
\section{Results and discussions}
In this section, we present and discuss the principal results obtained for the model. We begin with the noninteracting limit to establish the effects of quasiperiodic modulation and staggered magnetic ordering on the persistent-current profiles and their spin-resolved contributions. We then investigate the interacting regime to elucidate how Hubbard correlations modify these responses, with particular emphasis on the emergence of distinct current-enhancement regimes and the re-entrant behavior at finite interaction strength. We further examine the corresponding ground-state energy variations and Drude weight to gain additional insight into the underlying transport characteristics. Finally, we assess the reliability of the mean-field theory (MFT) by comparing its predictions with exact diagonalization (ED) and density-matrix renormalization group (DMRG) calculations for a finite system at half filling. Unless stated otherwise, all calculations are performed for a system of size $L=144$, with the staggered antiferromagnetic field fixed at $h_z=0.4$ and the reference hopping amplitude set to $t=1$. All energy parameters, including the quasiperiodic modulation strengths, hopping and the on-site Hubbard interaction, are expressed in electronvolts (eV), while the persistent charge and spin currents are reported in microamperes ($\mu\mathrm{A}$). The remaining model parameters and their corresponding values are specified alongside the relevant calculations and figures. These parameter choices are adopted as the default throughout the analysis unless explicitly stated otherwise.
\subsection{Non-interacting case}
We begin by examining the noninteracting limit ($U=0$) to establish how quasiperiodic modulation and magnetic ordering influence the persistent-current response. We first consider the nonmagnetic configuration as a reference case, followed by the ferromagnetic (FM) and antiferromagnetic (AFM) configurations to elucidate the role of different magnetic textures. By comparing the modulation-dependent charge and spin currents across these configurations, we identify how magnetic ordering modifies the current response and the relative contributions of the two spin channels. This analysis establishes the noninteracting baseline for understanding the additional effects of Hubbard correlations in the interacting regime.
\subsubsection{Nonmagnetic Configuration}
\label{subsec:nm}
\begin{figure}[t]
    \centering
    \includegraphics[width=8.5cm,height=4cm]{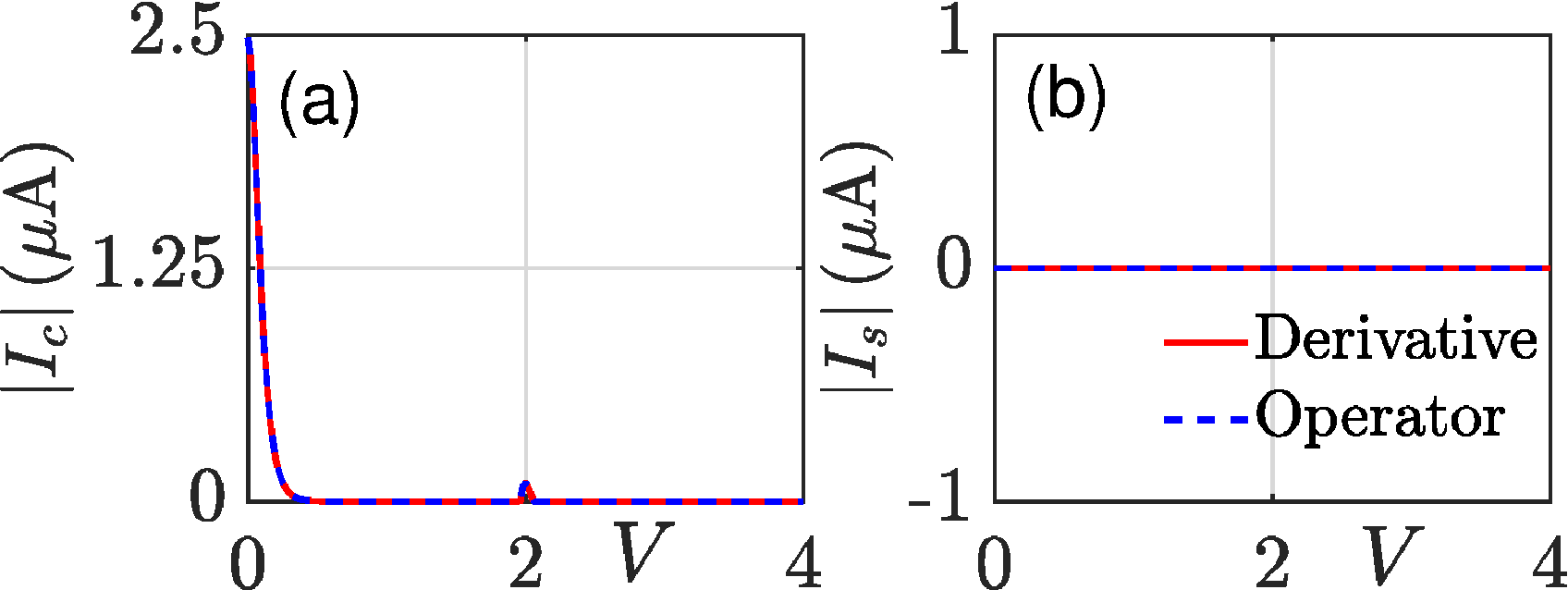}
    \caption{Modulation-strength dependence of the absolute persistent-current amplitudes for a nonmagnetic ring at half filling with $N=144$ sites, threaded by a flux $\phi=0.3\phi_0$ and with fixed hopping-modulation strength $W=0.2$. (a) Charge current $|I_c|$ and (b) spin current $|I_s|$ as functions of $V$.}
    \label{fig:nm}
\end{figure}
We first examine the persistent-current response of the nonmagnetic (NM) configuration, as shown in Fig.~\ref{fig:nm}. All calculations are performed for a ring comprising $N=144$ sites at half filling, with an Aharonov-Bohm flux $\phi=0.3$. The system is described by a generalized Aubry-Andr\'e-Harper (GAAH) modulation, while the strength of the hopping modulation is fixed at $W=0.2$ throughout the analysis. These parameters are kept unchanged when comparing the NM, ferromagnetic (FM), and antiferromagnetic (AFM) configurations, allowing the influence of magnetic ordering to be examined under identical conditions. Figure~\ref{fig:nm}(a) shows that the magnitude of the charge persistent current $|I_c|$ is largest in the weak-site-disorder limit, reaching approximately $2.5\,\mu\mathrm{A}$ near $V=0$. As $V$ increases, the current decreases rapidly and becomes negligibly small over most of the investigated interval. A very weak feature around $V\simeq 2$ is also visible, but its magnitude is substantially smaller than that of the clean-limit response. This behavior indicates that the site-energy modulation progressively suppresses the flux-sensitive electronic circulation. Although the GAAH potential is spatially correlated rather than random, its increasing amplitude modifies the distribution and spatial character of the electronic states, thereby reducing their contribution to the net persistent current. The spin-current response, presented in Fig.~\ref{fig:nm}(b), remains zero throughout the plotted range. In the absence of magnetic ordering and spin-dependent terms that distinguish the two spin sectors, the spin-resolved charge currents are equivalent and cancel in the spin-current difference, while adding in the charge-current sum. Consequently, the NM ring supports a substantial charge current primarily in the weak-disorder regime but exhibits no net spin persistent current. This combination of rapid charge-current suppression and vanishing spin response establishes the baseline against which the effects of FM and AFM ordering can be assessed.

\subsubsection{Ferromagnetic Configuration}
\label{subsec:fm}
\begin{figure}[t]
    \centering
    \includegraphics[width=8.5cm,height=4cm]{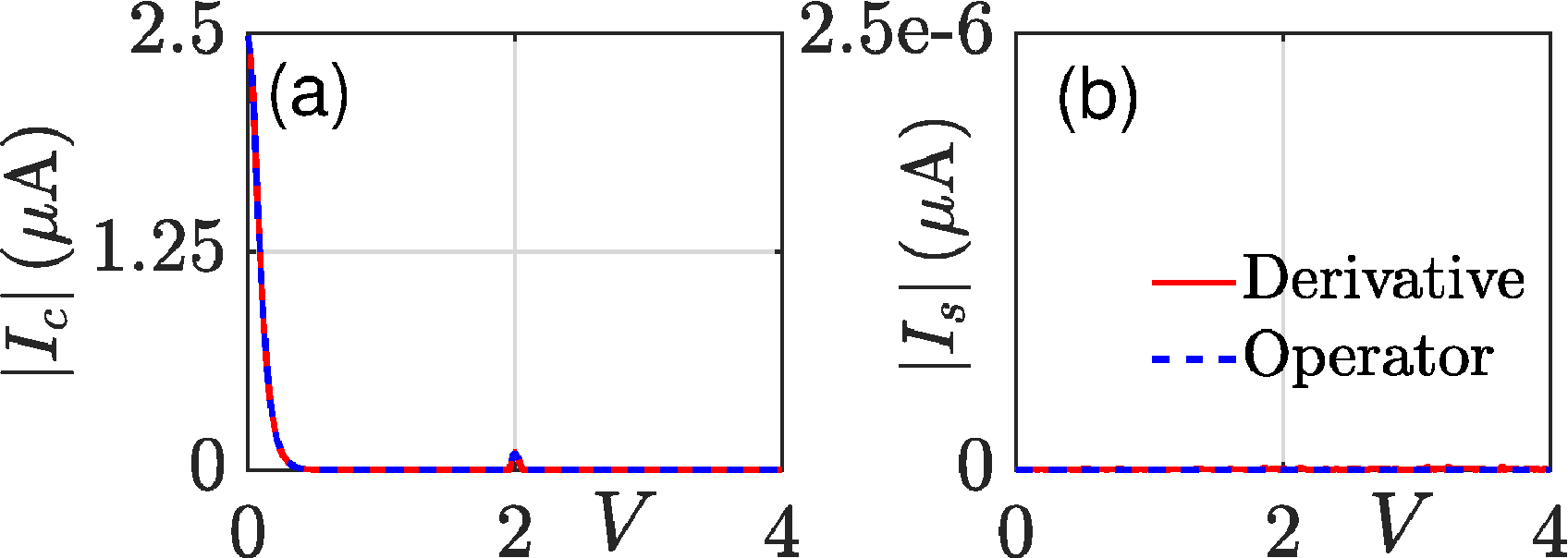}
    \caption{A plot analogous to Fig.~\ref{fig:nm} for the ferromagnetic configuration with $h_z=0.4$.}
    \label{fig:fm}
\end{figure}
Figure~\ref{fig:fm} presents the persistent currents for the ferromagnetic (FM) configuration, using the same system size $N=144$, half-filled occupation, flux $\phi=0.3$, and fixed hopping-modulation strength $W=0.2$. The only intended change from the NM calculation is the magnetic configuration, so the differences in the current profiles can be associated with the introduction of ferromagnetic ordering within the adopted model. As displayed in Fig.~\ref{fig:fm}(a), the charge current $|I_c|$ reaches approximately $2.5\,\mu\mathrm{A}$ near $V=0$ and then decreases steeply as the site-disorder strength increases. At larger values of $V$, the current is nearly suppressed, apart from a small residual feature around $V\simeq 2$. Thus, the FM configuration retains a strong charge-current response in the weak-disorder limit but does not sustain a comparable response across the finite-disorder regime. The persistence of this trend despite the introduction of magnetic ordering suggests that, for the chosen parameters, the GAAH site modulation remains the dominant factor controlling the overall charge-current magnitude. Figure~\ref{fig:fm}(b) shows that the spin current $|I_s|$ is extremely small, with a vertical scale of approximately $2.5\times10^{-6}\,\mu\mathrm{A}$. This is negligible compared with the charge-current scale. Ferromagnetic ordering distinguishes the spin sectors, but a nonzero spin polarization alone does not guarantee a large circulating spin current. The latter depends on the difference between the flux-induced responses of the two spin channels, which may remain small even when their individual contributions are finite. A comparison of Figs.~\ref{fig:nm} and~\ref{fig:fm} therefore shows that FM ordering does not qualitatively alter the dominant decreasing trend of the charge current and does not generate an appreciable spin-current response under the conditions studied.

\subsubsection{Antiferromagnetic Configuration Without Hopping Modulation}
\label{subsec:afm_W0}
\begin{figure}[t]
    \centering
    \includegraphics[width=8.5cm,height=4cm]{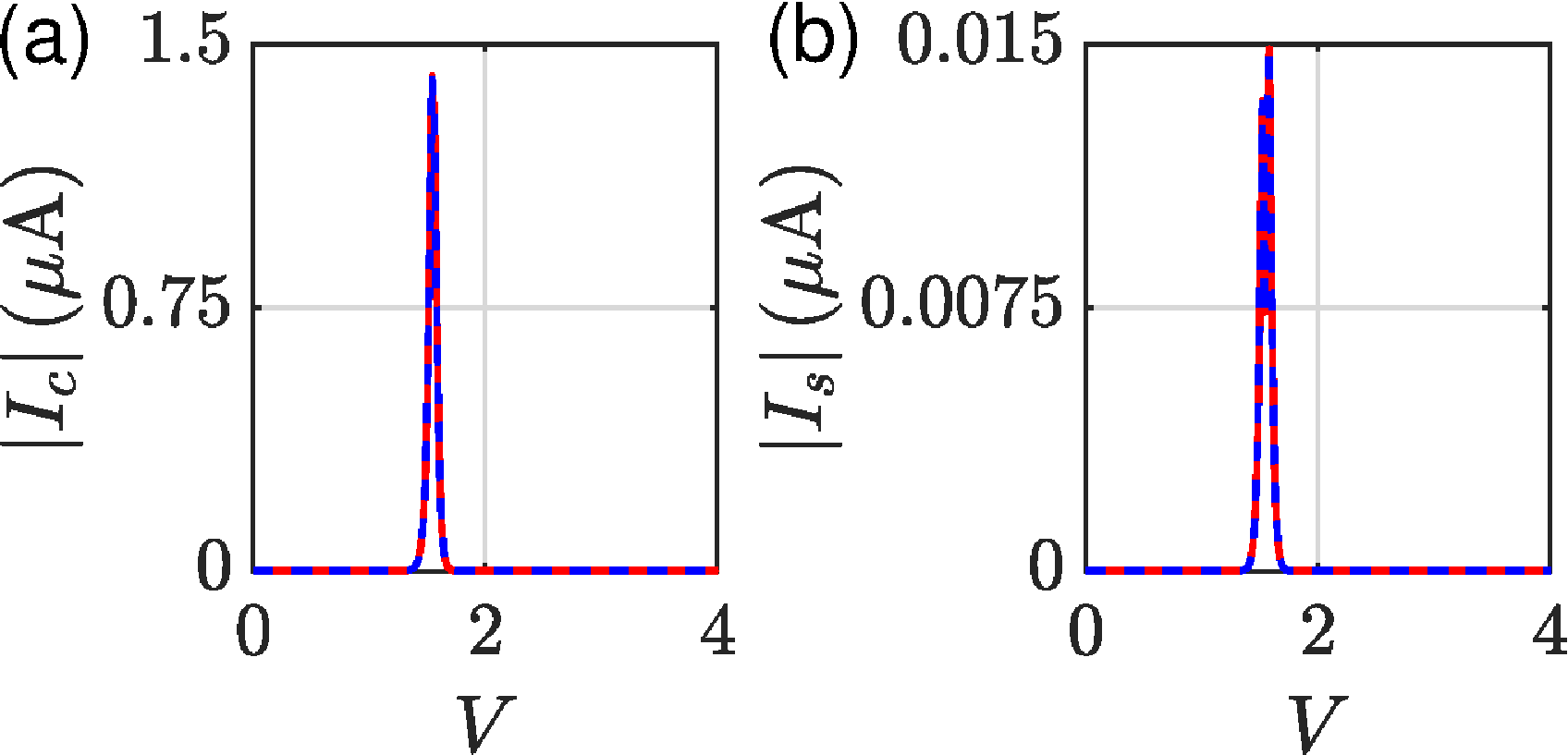}
    \caption{Variation of the persistent charge and spin currents for the antiferromagnetic configuration with $W=0$ and $h_z=0.4$. (a) Charge current $|I_c|$ and (b) spin current $|I_s|$ as functions of $V$, with all other parameters kept identical to those in Fig.~\ref{fig:nm}.}
    \label{fig:afm2}
\end{figure}
For the antiferromagnetic (AFM) configuration without hopping modulation, $W=0$, we investigate the persistent-current response of a ring comprising $N=144$ sites at half filling under an Aharonov-Bohm flux $\phi=0.3$, as shown in Fig.~\ref{fig:afm2}. The site-disorder strength $V$ is varied while all other parameters are held fixed. Figure~\ref{fig:afm2}(a) shows that the magnitude of the charge persistent current $|I_c|$ exhibits a sharp and well-defined peak near $V\simeq 1.3$, reaching approximately $1.4\,\mu\mathrm{A}$. Away from this narrow interval, the current rapidly decreases and remains negligibly small over the rest of the investigated range. This pronounced maximum indicates that the charge-current response is strongly dependent on the site-disorder strength, with appreciable circulation restricted to a narrow parameter window. The spin-current response, shown in Fig.~\ref{fig:afm2}(b), exhibits a peak at approximately the same value of $V$, suggesting that the charge and spin currents are sensitive to similar disorder-dependent changes in the electronic structure. However, the magnitude of the spin current is considerably smaller than that of the charge current: its maximum is only about $0.015\,\mu\mathrm{A}$, compared with approximately $1.4\,\mu\mathrm{A}$ for the charge current. Thus, despite the presence of antiferromagnetic ordering, the circulating response is predominantly charge dominated, and the spin current constitutes a much weaker contribution. The small spin-current magnitude indicates that the difference between the spin-resolved circulating currents remains limited, even near the disorder strength at which the charge current is maximal. The coincidence of the two maxima therefore reflects a common sensitivity to the electronic states in this narrow parameter region, rather than comparable magnitudes of charge and spin transport. Overall, the $W=0$ results reveal a sharply localized current-response window in the disorder parameter space, characterized by a pronounced charge-current peak and a substantially weaker spin-current peak. This behavior also establishes the baseline response of the AFM ring in the absence of hopping modulation.

\subsubsection{Antiferromagnetic Configuration at Finite Hopping Modulation}
\label{subsec:afm}
\begin{figure}[t]
    \centering
    \includegraphics[width=8.5cm,height=4cm]{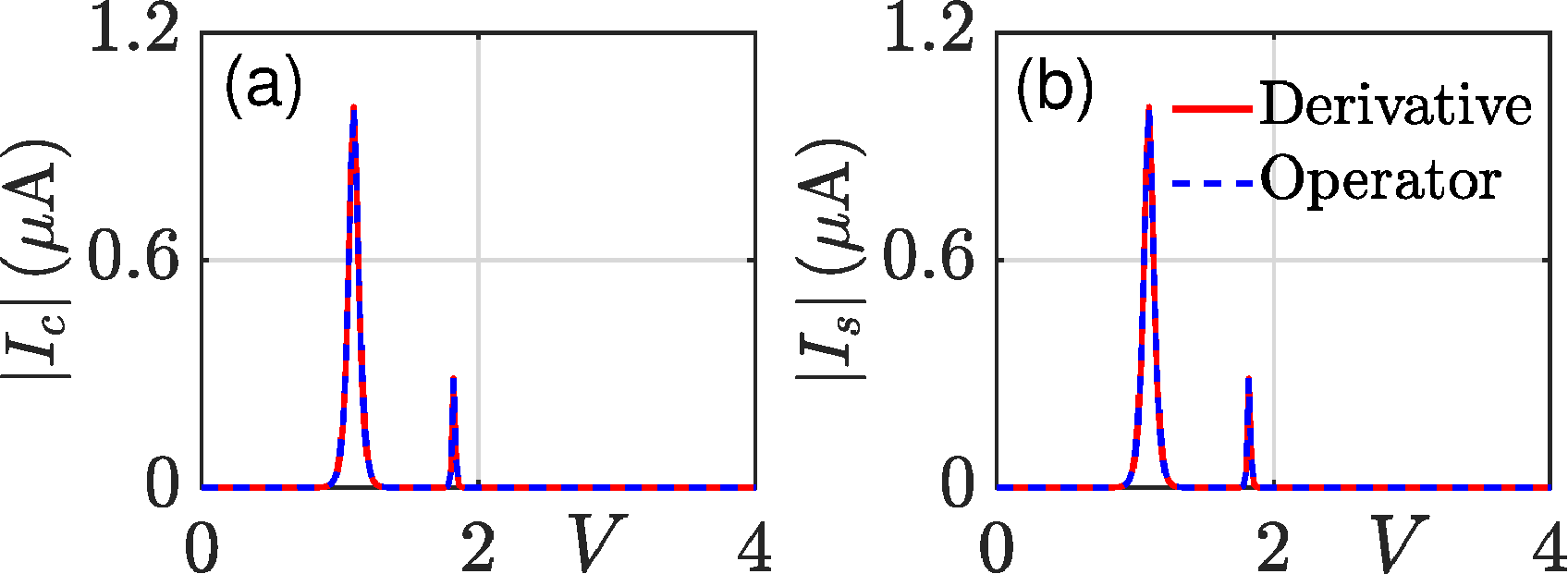}
    \caption{Dependence of the absolute persistent-current amplitudes on the site-disorder strength $V$ at a flux $\phi=0.3\phi_0$ and fixed hopping-modulation strength $W=0.2$ for a ring of $N=144$ sites under half filling at $h_z=0.4$. (a) Charge current $|I_c|$ and (b) spin current $|I_s|$ as functions of $V$.}
    \label{fig:afm}
\end{figure}
We now investigate the persistent-current response of the antiferromagnetic (AFM) ring in the presence of finite hopping modulation, as presented in Fig.~\ref{fig:afm}. The system consists of $N=144$ sites at half filling and is threaded by an Aharonov-Bohm flux $\phi=0.3$. The hopping-modulation strength is fixed at $W=0.2$, while the site-disorder strength $V$ is varied. Figure~\ref{fig:afm}(a) reveals a distinctly nonmonotonic dependence of the charge persistent current $|I_c|$ on $V$, characterized by two well-separated maxima. The first and more pronounced peak occurs near $V\simeq 1.1$, with a current magnitude of approximately $1\,\mu\mathrm{A}$, whereas a second peak emerges around $V\simeq 1.8$, reaching approximately $0.3\,\mu\mathrm{A}$. Between these maxima, the current undergoes substantial suppression, and its magnitude remains small over much of the remaining parameter range. The occurrence of two distinct current-enhancement windows indicates that the response cannot be described by a simple monotonic reduction of charge circulation with increasing site-disorder strength. Instead, appreciable current reappears at separated values of $V$, giving rise to a re-entrant-like structure in the current profile. The spin-current response in Fig.~\ref{fig:afm}(b) exhibits a related nonmonotonic dependence, with two features occurring approximately in the same disorder regions as the charge-current maxima. Nevertheless, the spin current remains substantially smaller in magnitude than the charge current, indicating that the enhancement of charge circulation is not accompanied by an equally strong spin-current response. The simultaneous occurrence of the peaks suggests that both current components are sensitive to the disorder-dependent reconstruction of the electronic states, although their magnitudes depend differently on the spin-resolved contributions. Physically, the finite hopping modulation introduces an additional spatially varying component into the electronic Hamiltonian, which competes with the GAAH site potential and the staggered magnetic texture. Their combined influence can modify the distribution, spatial character, and flux sensitivity of the electronic states as $V$ is varied. Consequently, the contributions of individual states to the persistent current may be suppressed over one parameter interval and become appreciable again in another, producing the observed two-peak structure. The AFM configuration therefore provides a means of controlling the charge-current response through the combined action of site disorder and hopping modulation, while also allowing a weaker spin-current response to emerge in the corresponding parameter regions. The two distinct maxima constitute evidence of a nonmonotonic, re-entrant-like current response in the finite-$W$ regime. However, the current profiles alone do not establish that the two peaks correspond to separate localization-delocalization transitions; such an interpretation would require independent support from localization measures, spectral properties, and system-size scaling.

\emph{We also examine the AFM configuration for the case in which the site-energy modulation is non-staggered, in contrast to the staggered arrangement considered in the present analysis. For a system comprising $N=144$ sites, both the charge and spin persistent currents are found to be negligibly small in the non-staggered case. This comparison highlights the sensitivity of the persistent-current response to the spatial arrangement of the site energies and motivates our focus on the staggered potential configuration in the following discussion.}

\subsubsection{Spin-Resolved Origin of the Re-entrant Current Response}
\label{subsec:pc_flux}
\begin{figure}[t]
    \centering
    \includegraphics[width=8.5cm,height=8cm]{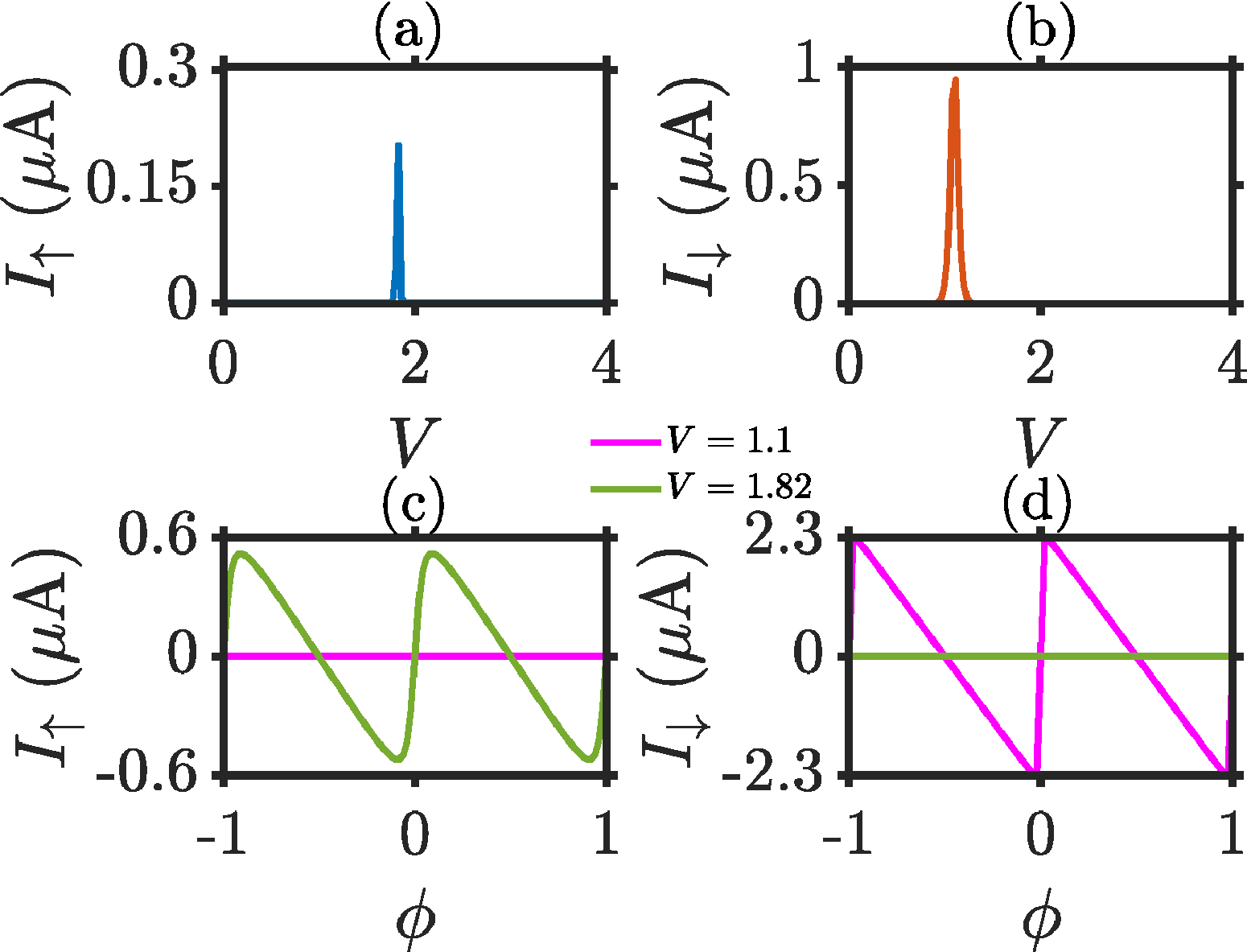}
    \caption{Spin-resolved persistent currents as functions of the modulation strength $V$ and the Aharonov-Bohm flux $\phi$ for a half-filled ring with $N=144$ sites. The first row displays the spin-up and spin-down currents versus $V$ at $\phi=0.3\phi_0$, while the second row shows their flux dependence at the two modulation strengths corresponding to the current peaks in the first row. The hopping-modulation strength is fixed at $W=0.2$ throughout.}
    \label{fig:pc_flux}
\end{figure}
To clarify the origin of the re-entrant current response in the antiferromagnetic (AFM) ring at $W=0.2$, we examine the spin-resolved persistent currents at two representative values of the site-potential modulation strength, corresponding to the primary and secondary current-enhancement regimes. The upper panels of Fig.~\ref{fig:pc_flux} show the spin-up current $I_{\uparrow}$ (left column) and spin-down current $I_{\downarrow}$ (right column) as functions of $V$, with $V=1.1$ and $V=1.82$ selected to represent the primary and secondary regimes, respectively. The lower panels display the corresponding flux dependence of these spin-resolved currents. A clear distinction emerges between the two regimes. At the primary enhancement point, $V=1.1$, the spin-down current is substantially larger in magnitude than the spin-up current, indicating that the current response is dominated by the spin-down channel, while the spin-up contribution is strongly suppressed. At the secondary enhancement point, $V=1.82$, the spin-up current becomes appreciable, whereas the spin-down current is comparatively suppressed. Thus, the dominant spin channel switches between the primary and secondary current-enhancement regimes. The flux-dependent profiles in the lower panels further illustrate this contrast: the two spin channels exhibit distinct amplitudes and flux dependences, reflecting their different contributions to the persistent circulation. These results show that the re-entrant response is accompanied by a redistribution of the spin-resolved currents, with the dominant contribution shifting from one spin channel to the other as $V$ is varied. This spin-selective switching provides a useful perspective on the distinct current configurations underlying the primary and secondary enhancement regimes in the AFM ring. From an application perspective, the switching of the dominant spin channel between the primary and secondary current-enhancement regimes suggests a possible mechanism for spin-selective current control in magnetic mesoscopic rings. By tuning the site-potential modulation strength $V$, the system can be driven between regimes in which the persistent current is dominated by opposite spin channels. Such tunability could be relevant to the design of spin-dependent current-control elements and reconfigurable spintronic architectures, where controlling spin-resolved circulation is desirable. Moreover, the contrasting responses of the two spin channels provide a means of modifying the balance between charge and spin currents through an external control parameter, without changing the underlying geometry. These observations establish a potential route toward modulation-controlled spin selectivity in an antiferromagnetic quantum ring.

\subsubsection{Bond-Resolved Persistent Current}
\label{subsec:bond_current}
\begin{figure}[t]
    \centering
    \includegraphics[width=8.5cm,height=4cm]{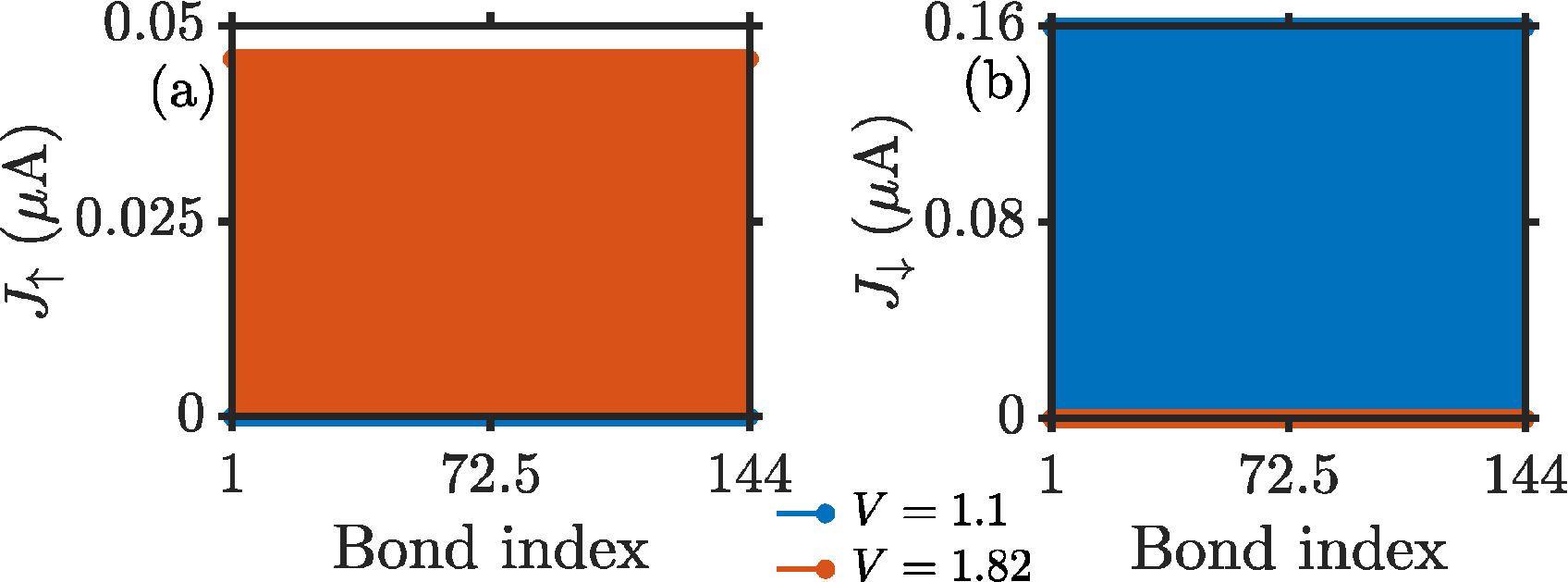}
    \caption{Spin-resolved bond-current distributions along the ring for the antiferromagnetic configuration. (a) Spin-up bond current $J_{\uparrow}$ and (b) spin-down bond current $J_{\downarrow}$ as functions of the bond index for the selected modulation strengths $V$. The remaining parameters are identical to those specified in the corresponding current calculation.}
    \label{fig:BOND}
\end{figure}
To gain further insight into the microscopic origin of the re-entrant current response, we investigate the spin-resolved bond-current distributions across the antiferromagnetic (AFM) ring at the two representative values of the site-potential modulation strength, $V=1.1$ and $V=1.82$, corresponding to the primary and secondary current-enhancement regimes, respectively. Figure~\ref{fig:BOND} displays the bond currents as functions of the bond index for the two spin channels. The left panel presents the spin-up bond current, while the right panel shows the spin-down bond current. A striking contrast emerges between the two parameter regimes. At the primary enhancement point, $V=1.1$, the spin-up current is nearly uniform across the entire ring, with a magnitude of approximately $0.047\,\mu\mathrm{A}$, whereas the spin-down contribution is strongly suppressed and remains close to zero. This indicates that the circulating current in this regime is predominantly carried by the spin-up channel. Upon increasing the modulation strength to the secondary enhancement point, $V=1.82$, the relative contributions of the two spin channels are substantially altered. The spin-up current becomes negligible, while the spin-down current rises to approximately $0.15\,\mu\mathrm{A}$ and remains nearly uniform along the ring. Consequently, the dominant contribution to the circulating current switches from the spin-up channel at the primary peak to the spin-down channel at the secondary peak. This pronounced spin-selective redistribution provides a spatially resolved signature of the re-entrant response and demonstrates that the secondary enhancement does not simply reproduce the current-carrying configuration found at the primary peak. Instead, changing $V$ reorganizes the relative contributions of the spin-resolved electronic states, leading to distinct current-carrying regimes. The nearly uniform bond-current profiles in both regimes are consistent with stationary circulation in a closed one-dimensional ring. In the absence of local sources or sinks, current continuity requires the net current entering and leaving each site to balance, so the steady-state bond current is spatially uniform. The substantial difference between the two profiles therefore arises primarily from the change in current magnitude and spin-channel dominance rather than from the formation of localized current paths along selected bonds. Microscopically, the site-potential modulation changes the electronic eigenstates and their sensitivity to the applied magnetic flux, thereby modifying the contributions of the occupied states to the persistent current. In the AFM configuration, the spin-dependent electronic structure can make these contributions different for the two spin channels. As $V$ is varied, the balance between the spin-up and spin-down contributions can consequently change, producing the observed switching of the dominant channel. The bond-resolved results thus complement the parameter-space maps and spin-resolved current analysis by identifying a clear distinction between the primary and secondary current-enhancement regimes. From an application perspective, this modulation-controlled switching suggests a possible route to spin-selective current control in magnetic mesoscopic rings. Nevertheless, establishing its usefulness for spintronic devices would require further investigation of spin polarization, robustness against parameter variations, and transport under an applied source-drain bias.

\subsubsection{Antiferromagnetic Configuration at Quarter Filling}
\label{subsec:afm_quarter}
\begin{figure}[t]
    \centering
    \includegraphics[width=8.5cm,height=4cm]{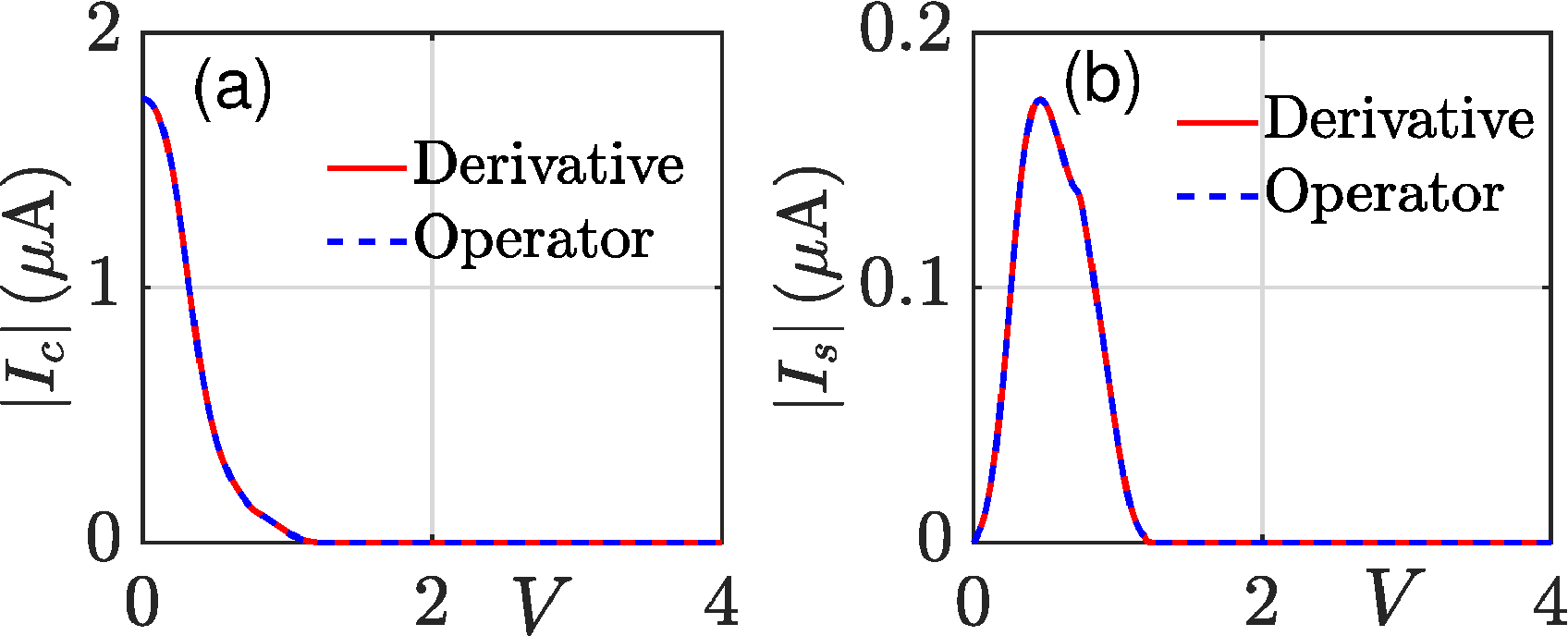}
    \caption{Persistent charge and spin currents as functions of the modulation strength $V$ for an antiferromagnetic ring at quarter filling. (a) Charge current $|I_c|$ and (b) spin current $|I_s|$. All other parameters are identical to those used in Fig.~\ref{fig:afm}.}
    \label{quarter1}
\end{figure}
We next examine the persistent-current response of the antiferromagnetic (AFM) ring at quarter filling in the presence of finite hopping modulation, $W=0.2$, as shown in Fig.~\ref{quarter1}. The charge and spin currents are evaluated as functions of the site-potential modulation strength $V$, with the results obtained using both the derivative and operator formulations. As shown in Fig.~\ref{quarter1}(a), the charge-current magnitude $|I_c|$ exhibits a pronounced peak at weak modulation, reaching approximately $1.5\,\mu\mathrm{A}$ near $V\simeq 0.4$. With increasing $V$, the current decreases rapidly and becomes negligibly small over the remainder of the displayed parameter range. The spin-current response in Fig.~\ref{quarter1}(b) displays a similar dependence on $V$, with a maximum of approximately $0.17\,\mu\mathrm{A}$ in the same weak-modulation regime, followed by a rapid suppression as $V$ increases. Thus, although the charge and spin currents exhibit peaks in approximately the same parameter region, the spin-current magnitude remains considerably smaller than the charge-current magnitude. The close agreement between the derivative and operator results provides a consistency check for the two current-evaluation approaches. Overall, the results indicate that at quarter filling, the persistent-current response of the AFM ring is concentrated in a narrow range of weak site-potential modulation and is strongly suppressed at larger $V$. This behavior reflects the sensitivity of flux-induced circulation to changes in the electronic states as the spatial modulation is increased. While the current profiles establish a pronounced suppression of both current components, identifying the underlying behavior as a localization transition would require additional evidence from suitable localization measures and system-size scaling.

\subsubsection{Flux Dependence of Persistent Currents at Half Filling}
\label{subsec:afm_flux}
\begin{figure}[t]
    \centering
    \includegraphics[width=8.5cm,height=4cm]{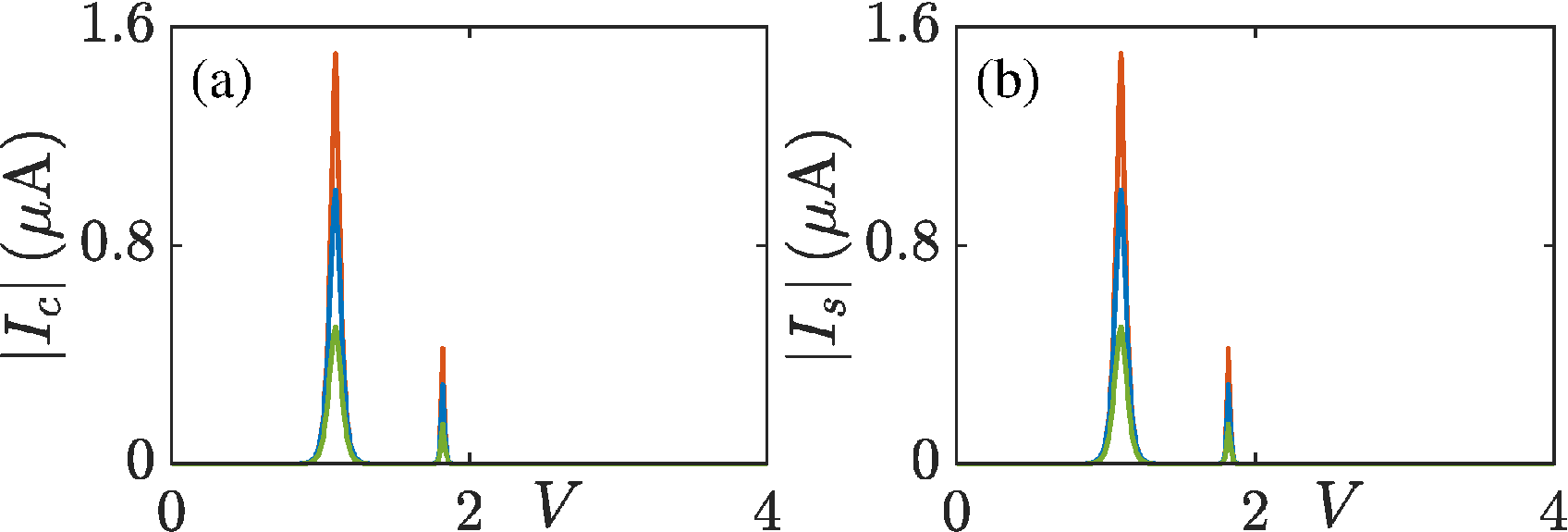}
    \caption{Dependence of the persistent charge and spin currents on the modulation strength $V$ for an antiferromagnetic ring with $N=144$ sites. (a) Charge current $|I_c|$ and (b) spin current $|I_s|$ for three Aharonov-Bohm flux values, $\phi=0.2\phi_0$ (orange), $\phi=0.3\phi_0$ (blue), and $\phi=0.4\phi_0$ (green). The hopping-modulation strength and staggered Zeeman field are fixed at $W=0.2$ and $h_z=0.4$, respectively.}
    \label{fig:AFM_HALF}
\end{figure}
We examine the robustness of the persistent-current response of the antiferromagnetic (AFM) ring at half filling against variations in the Aharonov-Bohm (AB) flux, with the hopping-modulation strength fixed at $W=0.2$. Figure~\ref{fig:AFM_HALF} shows the charge and spin currents as functions of the site-potential modulation strength $V$ for $\phi=0.2$, $0.3$, and $0.4$, represented by the orange, blue, and green curves, respectively. As shown in Fig.~\ref{fig:AFM_HALF}(a), the charge current exhibits a pronounced maximum near $V\simeq 1.1$, followed by a sharp suppression as $V$ increases. A weaker secondary peak emerges around $V\simeq 1.82$, indicating the reappearance of a finite current response after the intervening suppression. The spin current in Fig.~\ref{fig:AFM_HALF}(b) exhibits a qualitatively similar profile, with a dominant peak in the weak-modulation regime and a much smaller secondary feature at approximately the same value of $V$. Although the peak amplitudes vary with the applied flux, the overall current profiles retain the same qualitative features across all three flux values. In particular, the persistence of the secondary current-enhancement window following a pronounced suppression demonstrates that the nonmonotonic, re-entrant-like response is robust against the flux variations considered here. These observations indicate that the AB flux primarily modulates the magnitude of the persistent currents without qualitatively altering the locations of the principal current-enhancement windows. Thus, the re-entrant-like current response is not restricted to a particular flux value but persists over the investigated range of $\phi$. This robustness highlights the role of the interplay between the quasiperiodic site potential, hopping modulation, and AFM texture in shaping the current response, while the AB flux provides an additional means of tuning its amplitude. We emphasize that the persistence of these current-enhancement windows establishes a robust re-entrant-like feature in the transport response.

\subsubsection{Flux Dependence of Persistent Currents at Quarter Filling}
\label{subsec:afm_quarter_flux}
\begin{figure}[t]
    \centering
    \includegraphics[width=8.5cm,height=4cm]{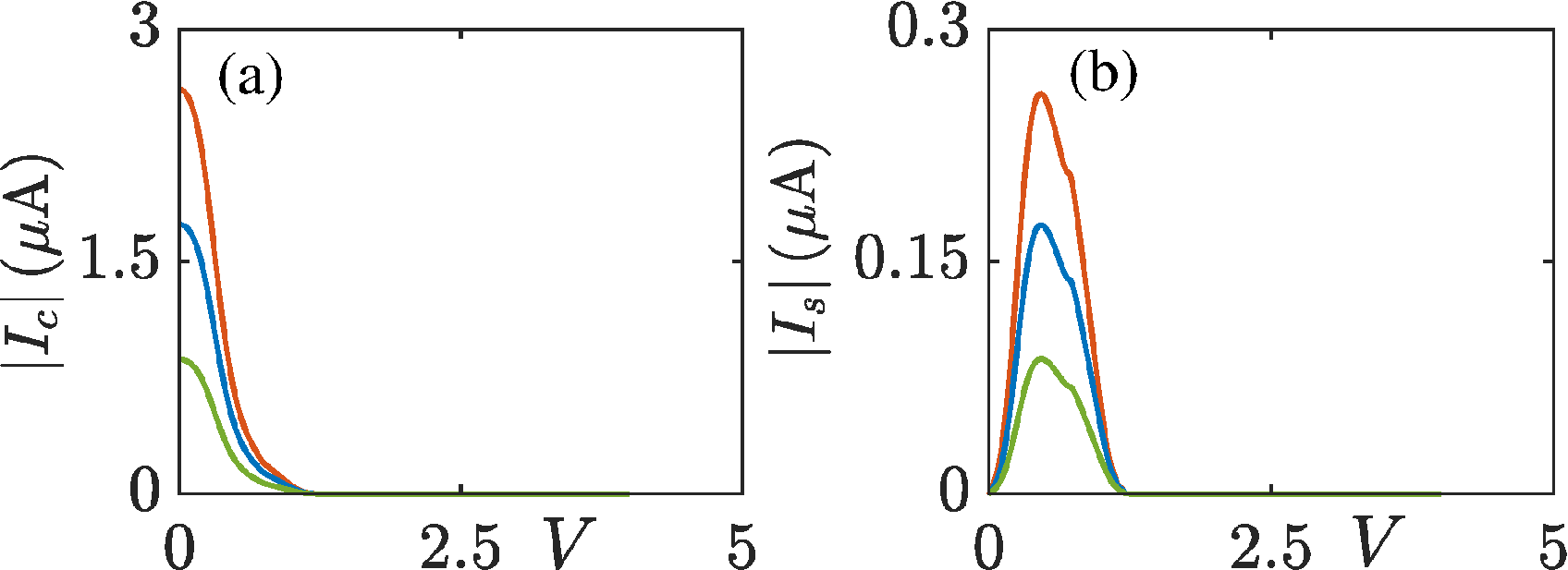}
    \caption{Persistent charge and spin currents as functions of the modulation strength $V$ for an antiferromagnetic ring at quarter filling. The plot follows the same parameter settings as Fig.~\ref{fig:AFM_HALF}, with the filling changed from half filling to quarter filling.}
    \label{fig:AFM_QUARTER}
\end{figure}
We further examine the robustness of the persistent-current response of the antiferromagnetic (AFM) ring at quarter filling for a finite hopping-modulation strength $W=0.2$, considering different values of the Aharonov-Bohm (AB) flux. Figure~\ref{fig:AFM_QUARTER} presents the charge and spin currents as functions of the site-potential modulation strength $V$ for three flux values. As shown in Fig.~\ref{fig:AFM_QUARTER}(a), the charge current is largest in the weak-modulation regime and decreases rapidly as $V$ increases. Although the peak magnitude and the rate of suppression depend on the applied flux, the overall trend remains qualitatively similar across the flux values considered. The spin-current response in Fig.~\ref{fig:AFM_QUARTER}(b) also exhibits a pronounced flux dependence, with its largest response concentrated in the weak-modulation regime. In contrast to the charge current, however, the spin-current amplitudes show more noticeable variation with flux, while becoming negligible at larger $V$. The persistence of the dominant current-response region across different flux values indicates that the qualitative behavior is robust against the flux variations considered, even though the magnitude of the charge and spin currents can change substantially. In particular, the rapid suppression of the currents with increasing $V$ remains a common feature, whereas the detailed peak amplitudes are flux dependent. These results demonstrate that the AB flux provides an effective means of tuning the magnitude of the persistent currents without fundamentally changing their overall dependence on the site-potential modulation. Unlike a clear re-entrant profile characterized by a secondary current-enhancement window after an intervening suppression, the present quarter-filling data primarily show a dominant low-$V$ response followed by rapid decay. Thus, the robustness here lies in the persistence of this qualitative trend rather than in a distinct re-entrant window.

\subsubsection{Extent of the Re-entrant Regime in the $(V,W)$ Plane}
\label{subsec:WV_half}
\begin{figure}[t]
    \centering
    \includegraphics[width=8.5cm,height=4cm]{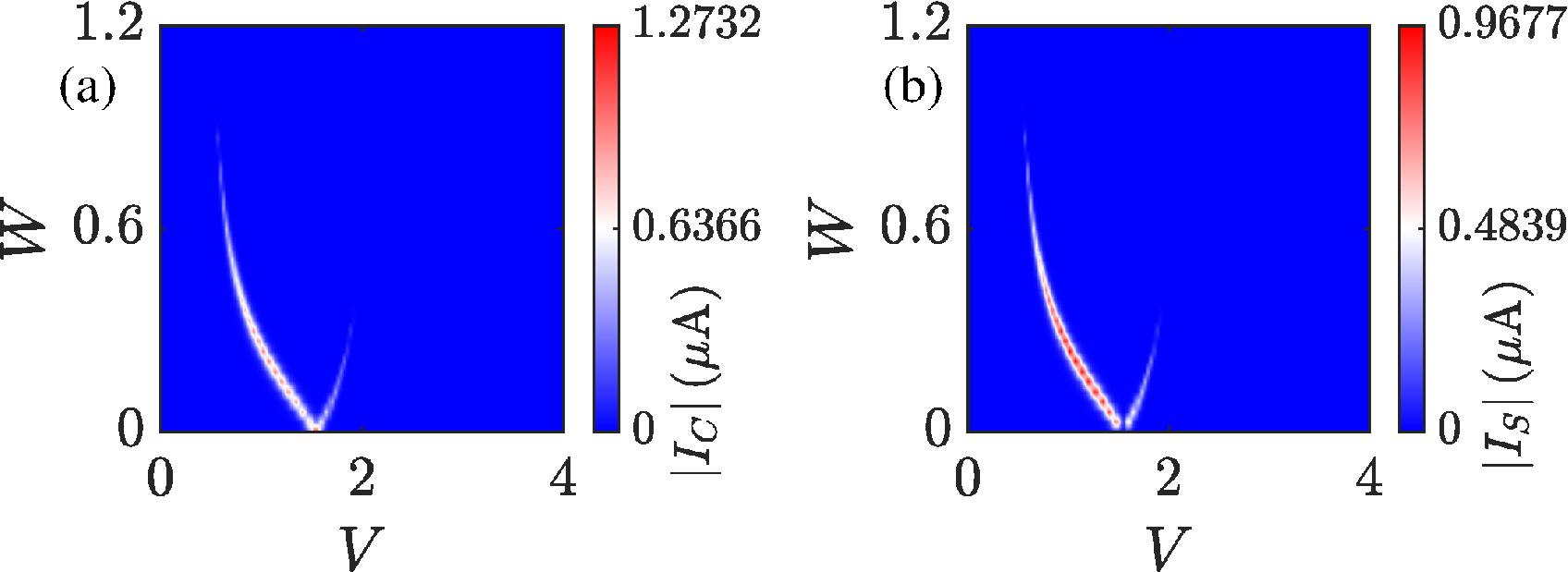}
    \caption{Heat maps of the persistent-current response in the $(W,V)$ parameter plane for an antiferromagnetic ring with $N=144$ sites at half filling. (a) Absolute charge current $|I_c|$ and (b) absolute spin current $|I_s|$ as functions of the hopping-modulation strength $W$ and site-potential modulation strength $V$. The Aharonov-Bohm flux and staggered Zeeman field are fixed at $\phi=0.3\phi_0$ and $h_z=0.4$, respectively.}
    \label{fig:WV_HALF}
\end{figure}
To characterize the evolution of the persistent currents across the parameter space at half filling, we map the charge- and spin-current magnitudes in the $(V,W)$ plane, as shown in Fig.~\ref{fig:WV_HALF}. The resulting maps reveal a strongly nonuniform current response, with pronounced enhancements confined to narrow regions of the site-potential modulation strength $V$ and hopping-modulation amplitude $W$. In both panels, the dominant current-enhancement feature follows a curved trajectory, indicating that the value of $V$ associated with an appreciable current response shifts as $W$ varies. Importantly, the secondary current-enhancement regime associated with the re-entrant-like response appears only over a restricted range of $W$, rather than persisting throughout the entire parameter space. As $W$ moves beyond this interval, the secondary feature becomes progressively suppressed and eventually loses its prominence, demonstrating that the re-entrant current response is conditional on the strength of the hopping modulation. Although the charge and spin currents display broadly similar structures in the $(V,W)$ plane, their magnitudes and intensity distributions differ, reflecting their distinct spin-resolved contributions. These maps therefore identify a finite window of hopping-modulation strength within which the secondary re-entrant-like current regime can emerge. The continuous parameter-space sweep highlights the importance of jointly tuning $V$ and $W$ to access this regime and demonstrates that the qualitative current response can change substantially across the parameter plane. We stress that the re-entrant feature discussed here refers to the recovery of the persistent-current response after its suppression.

\subsubsection{Extent of the Re-entrant Regime in the $(V,W)$ Plane at Quarter Filling}
\label{subsec:WV_quarter}
\begin{figure}[t]
    \centering
    \includegraphics[width=8.5cm,height=4cm]{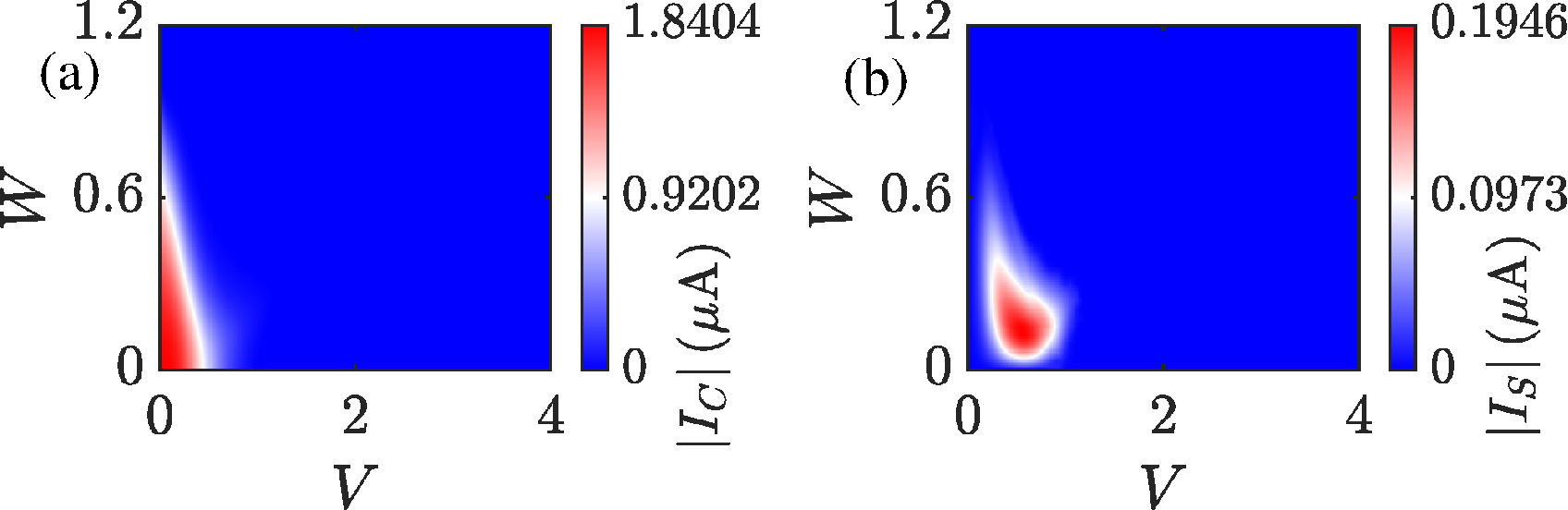}
    \caption{Corresponding $(W,V)$ current maps at quarter filling for the antiferromagnetic ring, with $\phi=0.3\phi_0$ and $h_z=0.4$. Panels (a) and (b) show the absolute charge and spin currents, respectively.}
    \label{fig:WV_QUARTER}
\end{figure}
To examine the parameter-space dependence of the persistent currents at quarter filling, we map the charge- and spin-current magnitudes in the $(V,W)$ plane, as shown in Fig.~\ref{fig:WV_QUARTER}. The charge-current map in Fig.~\ref{fig:WV_QUARTER}(a) exhibits a pronounced enhancement concentrated near weak site-potential modulation, with the response rapidly diminishing as $V$ increases. The spin-current distribution in Fig.~\ref{fig:WV_QUARTER}(b) displays a more localized enhancement at finite $V$ and relatively weak hopping modulation, followed by strong suppression away from this region. These distinct intensity distributions demonstrate that the charge and spin currents respond differently to the combined variation of $V$ and $W$. In contrast to a secondary enhancement that persists across an extended region of parameter space, the quarter-filled maps show that appreciable current responses are confined to restricted parameter windows. In particular, the spin-current enhancement is localized within a narrow region of the $(V,W)$ plane, indicating that its occurrence is sensitive to the precise combination of site-potential and hopping modulations. The current maps therefore highlight the filling dependence of the parameter-space structure and show that a re-entrant-like response, when identified through a secondary current-enhancement window, need not extend across the full range of hopping-modulation strengths. Rather, the relevant enhancement features are restricted to particular regions of the $(V,W)$ plane.

\subsubsection{Evolution of Persistent Currents with Staggered Magnetic Field at Half Filling}
\label{subsec:hzv_half}
\begin{figure}[t]
    \centering
    \includegraphics[width=8.5cm,height=4cm]{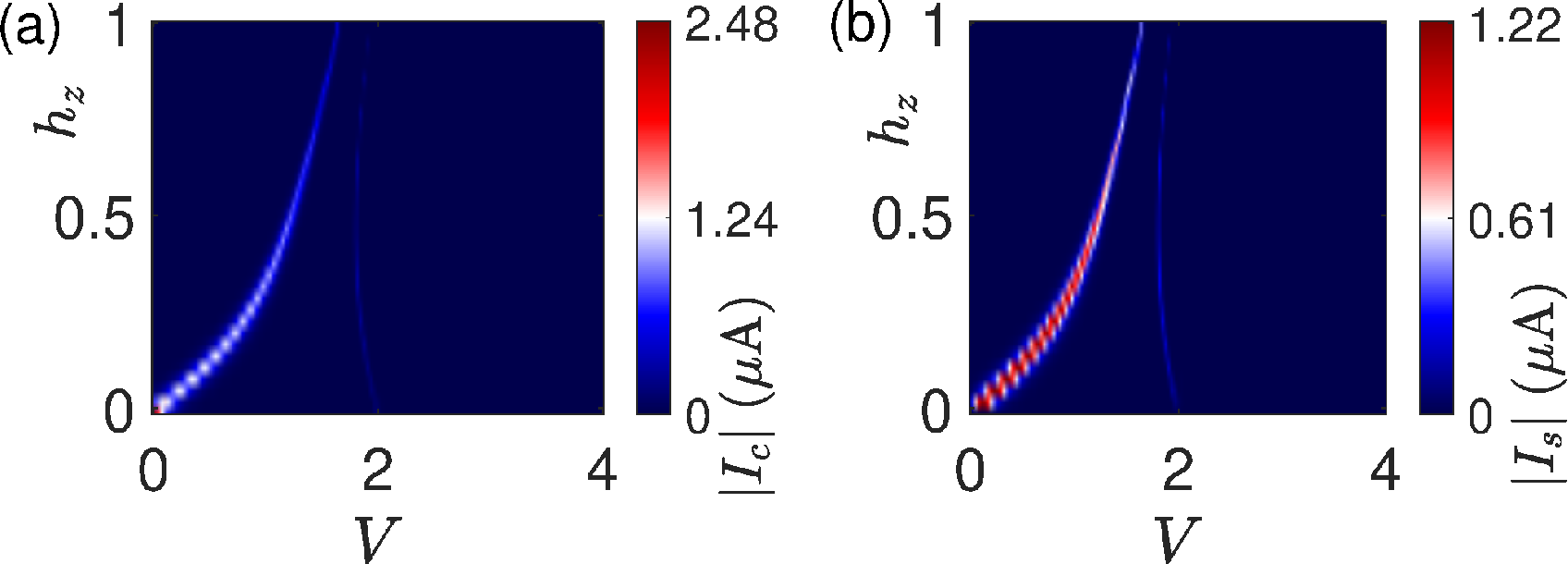}
    \caption{Heat maps of the persistent-current response in the $(h_z,V)$ parameter plane for an antiferromagnetic ring at half filling with $N=144$ sites. Panels (a) and (b) display the absolute charge current $|I_c|$ and spin current $|I_s|$, respectively, at fixed $W=0.2$ and $\phi=0.3\phi_0$.}
    \label{fig:HZV_HALF}
\end{figure}
We next investigate the dependence of the persistent currents on the staggered magnetic-field strength $h_z$ at half filling, with the hopping-modulation amplitude fixed at $W=0.2$. Figure~\ref{fig:HZV_HALF} presents the charge- and spin-current magnitudes in the $(V,h_z)$ parameter plane. The charge-current map in Fig.~\ref{fig:HZV_HALF}(a) reveals a pronounced, ridge-like enhancement that follows a curved trajectory as $h_z$ increases. For a given field strength, the dominant current response is concentrated around a specific range of $V$, with the peak position shifting systematically as $h_z$ varies. Importantly, the current enhancement remains organized around a single dominant peak along the $h_z$ sweep, rather than developing the distinct secondary enhancement window observed in the re-entrant-like response in other parameter sweeps. The spin-current map in Fig.~\ref{fig:HZV_HALF}(b) exhibits a corresponding dominant enhancement ridge, although its intensity distribution differs from that of the charge current and additional weaker features are visible near the principal ridge. Overall, the maps show that varying $h_z$ primarily shifts the location of the dominant current-enhancement region in the $(V,h_z)$ plane, while the response remains concentrated around a single principal peak. Thus, within the parameter range considered, the staggered magnetic field controls the position and magnitude of the persistent-current enhancement without producing a clear, separate secondary re-entrant window. These results highlight the distinct role of the magnetic texture in shaping the current response of the half-filled ring.

\subsubsection{Staggered Magnetic-Field Dependence at Quarter Filling}
\label{subsec:hzv_quarter}
\begin{figure}
    \centering
    \includegraphics[width=8.5cm,height=4cm]{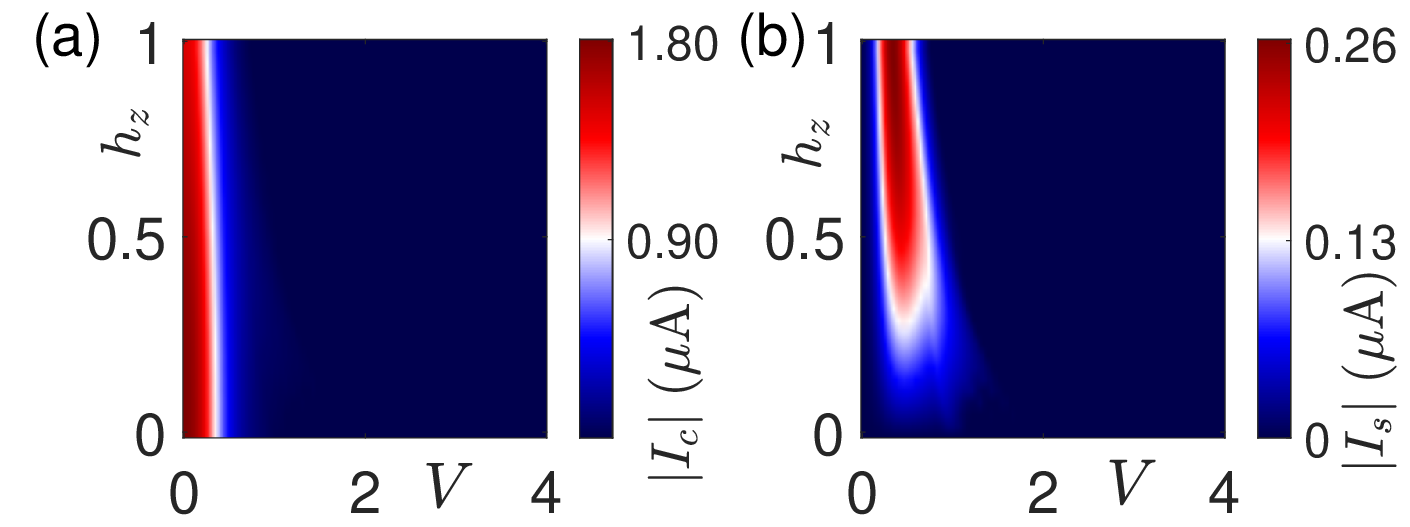}
    \caption{Quarter-filled counterpart of the $(h_z,V)$ current maps, with the hopping-modulation strength fixed at $W=0.2$ and the flux at $\phi=0.3\phi_0$. Panels (a) and (b) show the absolute charge current $|I_c|$ and spin current $|I_s|$, respectively, for $N=144$ sites.}
    \label{fig:HZV_QUARTER}
\end{figure}
We examine the persistent-current response of the antiferromagnetic ring at quarter filling in the $(V,h_z)$ parameter space, with the hopping-modulation strength fixed at $W=0.2$. Figure~\ref{fig:HZV_QUARTER} shows the charge- and spin-current magnitudes as functions of the site-potential modulation strength $V$ and the staggered magnetic-field amplitude $h_z$. The charge-current map in Fig.~\ref{fig:HZV_QUARTER}(a) exhibits a narrow, pronounced enhancement confined to the weak-$V$ region. This feature extends across the displayed range of $h_z$, while the current is strongly suppressed as $V$ increases, indicating that the dominant charge-current response remains concentrated near small site-potential modulation rather than shifting substantially with the magnetic field. The spin-current map in Fig.~\ref{fig:HZV_QUARTER}(b), by contrast, displays a more structured distribution. Its dominant enhancement also occurs at weak $V$, but the intensity broadens toward finite $V$ at lower and intermediate values of $h_z$, producing a localized region of enhanced spin current. As $h_z$ increases, this broader feature becomes less prominent and the response becomes concentrated closer to the low-$V$ region. Thus, unlike the charge current, the spin current exhibits a more pronounced dependence on the combined variation of $V$ and $h_z$. Importantly, neither map displays a clearly separated secondary enhancement window following an intervening region of suppressed current. The quarter-filled response is therefore characterized primarily by a dominant low-$V$ current-enhancement region, with the staggered magnetic field affecting the distribution most visibly in the spin-current map. These results underscore the distinct sensitivities of charge and spin circulation to the magnetic texture at quarter filling.

\subsubsection{Spin-Resolved Eigenstate properties}
\label{subsec:IPR}
\begin{figure}[t]
    \centering
    \includegraphics[width=8.5cm,height=7cm]{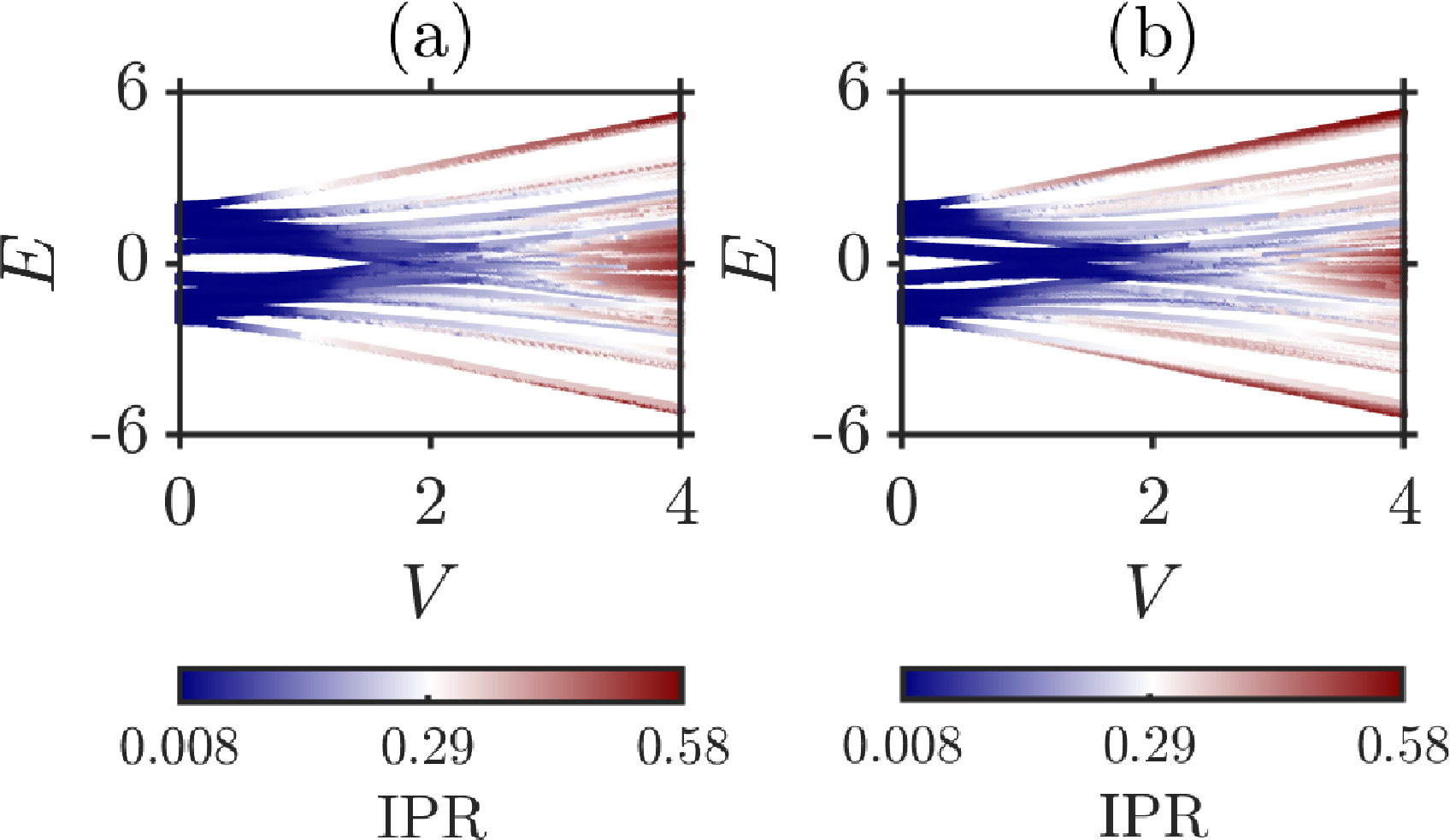}
    \caption{Energy spectrum of the antiferromagnetic ring as a function of the site-potential modulation strength $V$, with eigenstates color-coded according to their inverse participation ratio (IPR). Results are shown for $N=144$ sites at half filling, with $W=0.2$, $\phi=0.3\phi_0$, and $h_z=0.4$.}
    \label{fig:IPR}
\end{figure}
Figure~\ref{fig:IPR} presents the spin-resolved inverse participation ratio (IPR) in the energy-modulation-strength $(E,V)$ plane for the antiferromagnetic ring with $W=0.2$, where the left and right panels correspond to the spin-up and spin-down sectors, respectively. The seismic colormap provides a visual representation of the spatial character of the eigenstates: blue shades indicate smaller IPR values associated with more spatially extended states, red shades represent larger IPR values corresponding to stronger localization, and white shades denote intermediate IPR values. A prominent feature of the spectra is the evolution of the gap around zero energy as $V$ increases. In particular, the central gap narrows and tends to close near $V=1.1$ in Fig.~\ref{fig:IPR}(b) and $V=1.82$ in Fig.~\ref{fig:IPR}(a), corresponding to the primary and secondary peaks in the persistent-current response, respectively. The gap narrowing indicates that eigenstates approach the zero-energy region, increasing the availability of states near the Fermi energy when the system is considered at half filling. Importantly, the IPR distributions also reveal relatively small values for states near zero energy in the vicinity of these two modulation strengths. The associated blue-shaded regions indicate that these states are comparatively more extended than states characterized by larger IPR values. This combination of central-gap narrowing and reduced IPR provides a physically plausible explanation for the enhancement of the persistent current at both characteristic values of $V$. The narrowing of the gap brings additional states into the vicinity of the Fermi energy, while their more extended spatial character can facilitate coherent circulation around the ring. Since the persistent current is governed by the sensitivity of the occupied energy levels to the applied Aharonov-Bohm flux, such states can contribute appreciably when their energies respond strongly to changes in the flux. The two current maxima can therefore be understood as distinct parameter regimes in which the near-zero-energy spectrum and the spatial character of the eigenstates become favorable for enhanced circulation. Crucially, however, the spin-resolved bond-current analysis reveals that the dominant contribution changes between these regimes: the primary peak at $V=1.1$ is dominated by the spin-up channel, whereas the secondary peak at $V=1.82$ is dominated by the spin-down channel. Thus, although both enhancements are associated with gap narrowing and relatively extended states near zero energy, they are not identical current-carrying configurations. Instead, the spin-resolved electronic structure determines which channel predominantly contributes to the circulating current in each regime. The secondary peak consequently represents a re-entrant enhancement of the persistent current accompanied by a reversal of the dominant spin contribution. These findings establish a connection between the evolution of the central spectral gap, the spatial extent of near-zero-energy eigenstates, and the spin-selective nature of the re-entrant current response. A definitive identification of the individual states responsible for the enhancement would require a state-resolved evaluation of their IPRs, occupations, and flux derivatives.

\subsubsection{Effect of electron filling}
\begin{figure}[t]
    \centering
    \includegraphics[width=8.5cm,height=4cm]{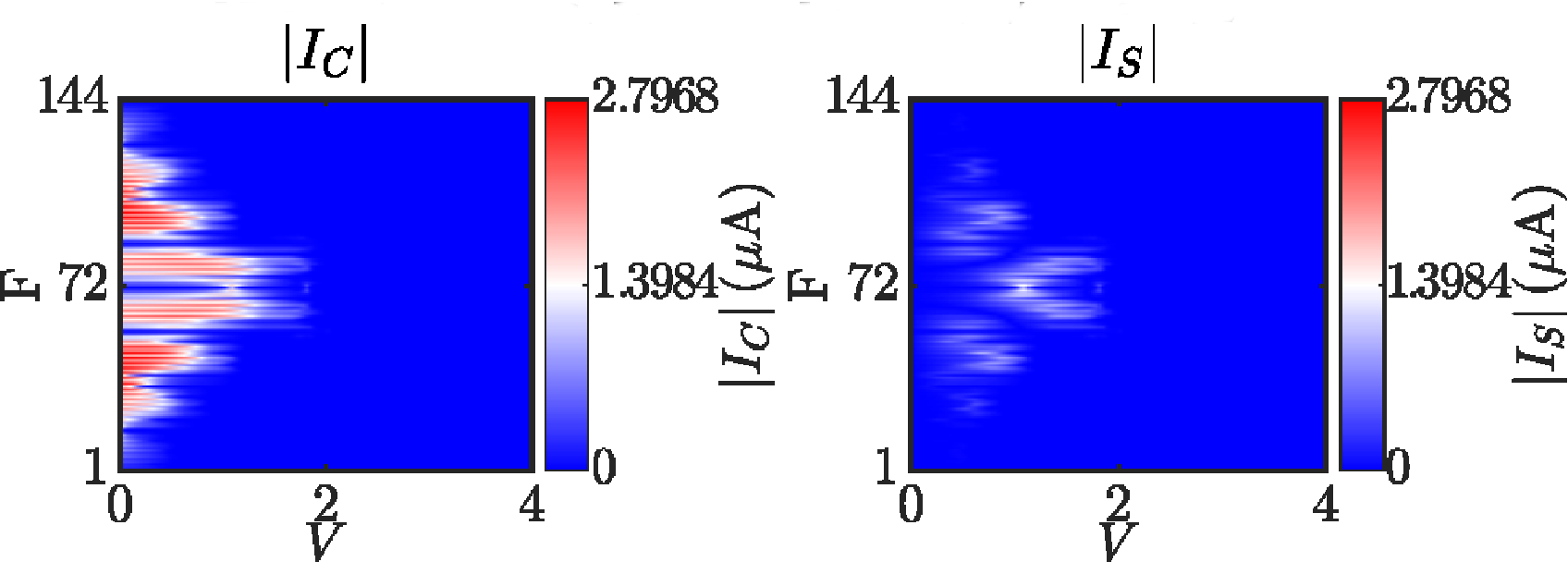}
    \caption{Filling-resolved distribution of the absolute charge and spin currents as functions of the modulation strength $V$. The left and right panels show the absolute charge current $|I_C|$ and absolute spin current $|I_S|$, respectively, for different electron fillings. The vertical axis represents the electron-filling index, $F$, ranging from $1$ to $144$, while the horizontal axis denotes the modulation strength $V$, varied from $0$ to $4$. The color maps represent the corresponding current magnitudes, with blue indicating vanishing or relatively small currents and red indicating larger current magnitudes. The color bars quantify $|I_C|$ and $|I_S|$ in $\mu\mathrm{A}$.}
    \label{fig:filling}
\end{figure}
Figure~\ref{fig:filling} illustrates the evolution of the absolute charge current $|I_C|$ and absolute spin current $|I_S|$ as functions of the modulation strength $V$ and electron-filling index $F$. The filling index varies from $F=1$ to $F=144$, while the modulation strength spans the interval $0\leq V\leq 4$. The color maps provide a two-dimensional representation of the current magnitudes, allowing the filling-dependent transport response to be examined across the parameter space.

The charge-current map in Fig.~\ref{fig:filling}(a) exhibits pronounced regions of enhanced current at selected filling indices, particularly at relatively weak modulation strengths. As $V$ increases, the charge current becomes strongly suppressed over most of the parameter space, as indicated by the predominantly blue regions. Nevertheless, several localized bright regions persist at intermediate modulation strengths, demonstrating that the current response is not uniformly suppressed with increasing $V$. Instead, the charge transport depends sensitively on the interplay between the modulation strength and electron filling.

The spin-current distribution in Fig.~\ref{fig:filling}(b) displays a qualitatively similar but comparatively weaker response. Enhanced spin-current regions occur only over restricted ranges of filling and modulation strength, while the current remains small across much of the parameter space. This contrast between the two panels indicates that the charge and spin currents exhibit distinct filling-dependent responses to the modulation. In particular, the regions of enhanced charge current do not necessarily coincide with those exhibiting enhanced spin current, suggesting that the two transport channels respond differently to changes in the system parameters.

Overall, Fig.~\ref{fig:filling} highlights the importance of electron filling in controlling charge and spin transport. The nonuniform current distributions indicate that the modulation strength alone does not determine the transport response; the filling-dependent electronic structure also plays an important role. Identifying the physical origin of the enhanced-current regions would require a complementary analysis of the energy-resolved spectrum, transmission properties, and current-carrying states.

\subsection{Effect of onsite Hubbard interaction}
Having established the role of quasiperiodic modulation and magnetic ordering in the noninteracting regime, we now turn to the effects of on-site Hubbard interactions. We investigate how increasing the interaction strength modifies the persistent charge and spin currents for different modulation strengths, with particular emphasis on the distinct responses of the two spin channels. By examining the current profiles alongside the corresponding ground-state energy variations and Drude weight, we explore how the interplay between quasiperiodicity, antiferromagnetic ordering, and Hubbard correlations governs the evolution of persistent currents.
\subsubsection{Hubbard Interaction Dependence of Charge and Spin Currents at Half Filling}
\begin{figure}[t]
    \centering
    \includegraphics[width=8.5cm,height=4cm]{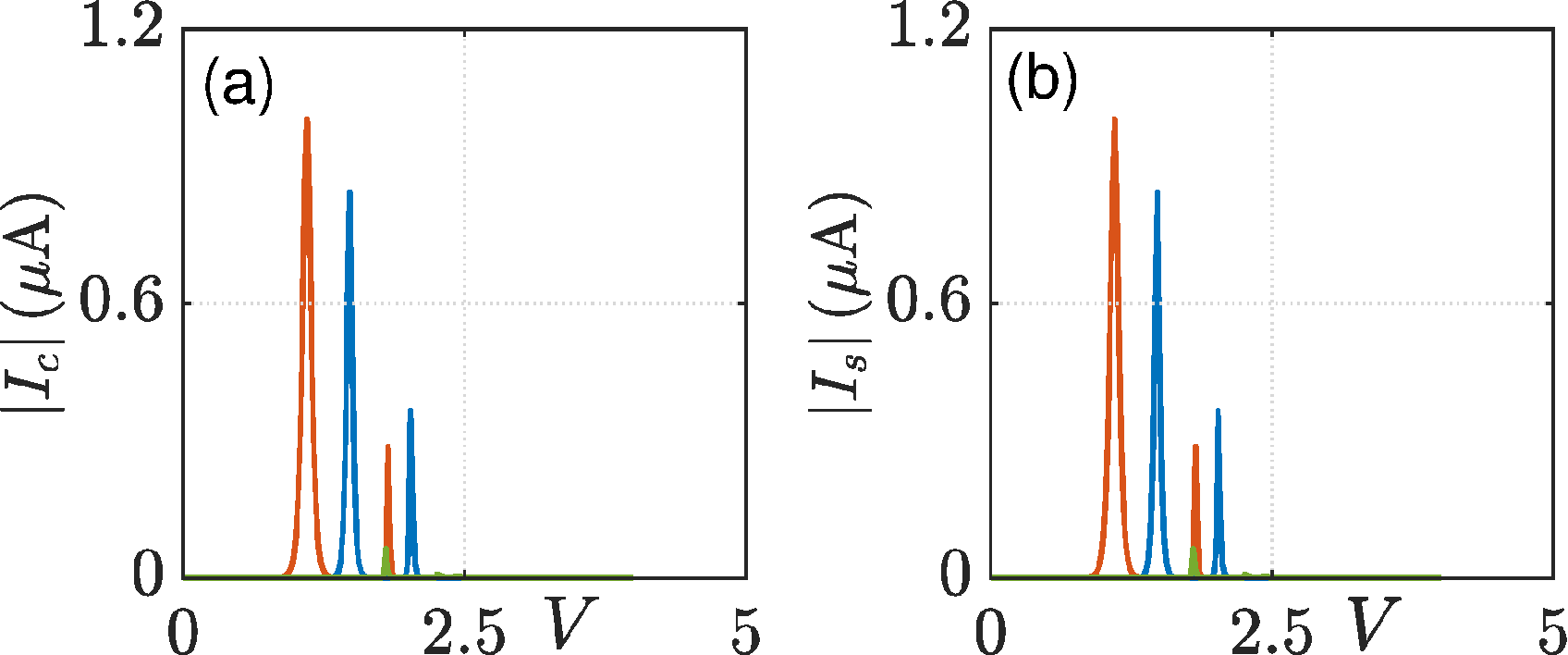}
    \caption{Dependence of the absolute persistent-current amplitudes on the modulation strength $V$ for an antiferromagnetic ring with $N=144$ sites. (a) Charge current $|I_c|$ and (b) spin current $|I_s|$ for Hubbard interaction strengths $U=0$, $1$, and $2$, represented by orange, blue, and green curves, respectively. The hopping-modulation strength, Aharonov-Bohm flux, and staggered Zeeman field are fixed at $W=0.2$, $\phi=0.3\phi_0$, and $h_z=0.4$, respectively.}
    \label{fig:U_HALF}
\end{figure}

To investigate the influence of the on-site Hubbard interaction on persistent-current transport at half filling, we compare the charge and spin currents for $U=0$, $1$, and $2$, represented by the red, blue, and green curves, respectively, in Fig.~\ref{fig:U_HALF}. The results reveal a contrasting evolution of the primary and secondary current peaks with increasing interaction strength. While the height of the primary peak decreases progressively with $U$, indicating a suppression of the dominant current response, the secondary peak appears to increase up to a certain interaction strength, suggesting that moderate electron-electron correlations can enhance the current response in this parameter regime. This opposing behavior indicates that the Hubbard interaction does not uniformly suppress persistent currents but instead modifies the relative contributions of the electronic states associated with different peak regions. Physically, the on-site repulsion changes the electronic spectrum and the spin-resolved character of the current-carrying states, thereby altering their sensitivity to the applied Aharonov-Bohm flux. Consequently, states contributing strongly to the primary peak may become less effective in carrying current as $U$ increases, while the contribution from states associated with the secondary peak may initially be enhanced. The apparent increase of the secondary peak over the investigated range suggests a nonmonotonic interaction dependence, although its behavior at stronger interactions requires further examination. These observations highlight the possibility of tuning the relative strengths of distinct persistent-current features through electron correlations, motivating a closer investigation of the interaction-dependent electronic states in the following analysis.

\subsubsection{Re-entrant Persistent-Current Response Induced by Hubbard Interaction}
\label{subsec:U_dependence_peaks}
\begin{figure}[t]
    \centering
    \includegraphics[width=8.5cm,height=8cm]{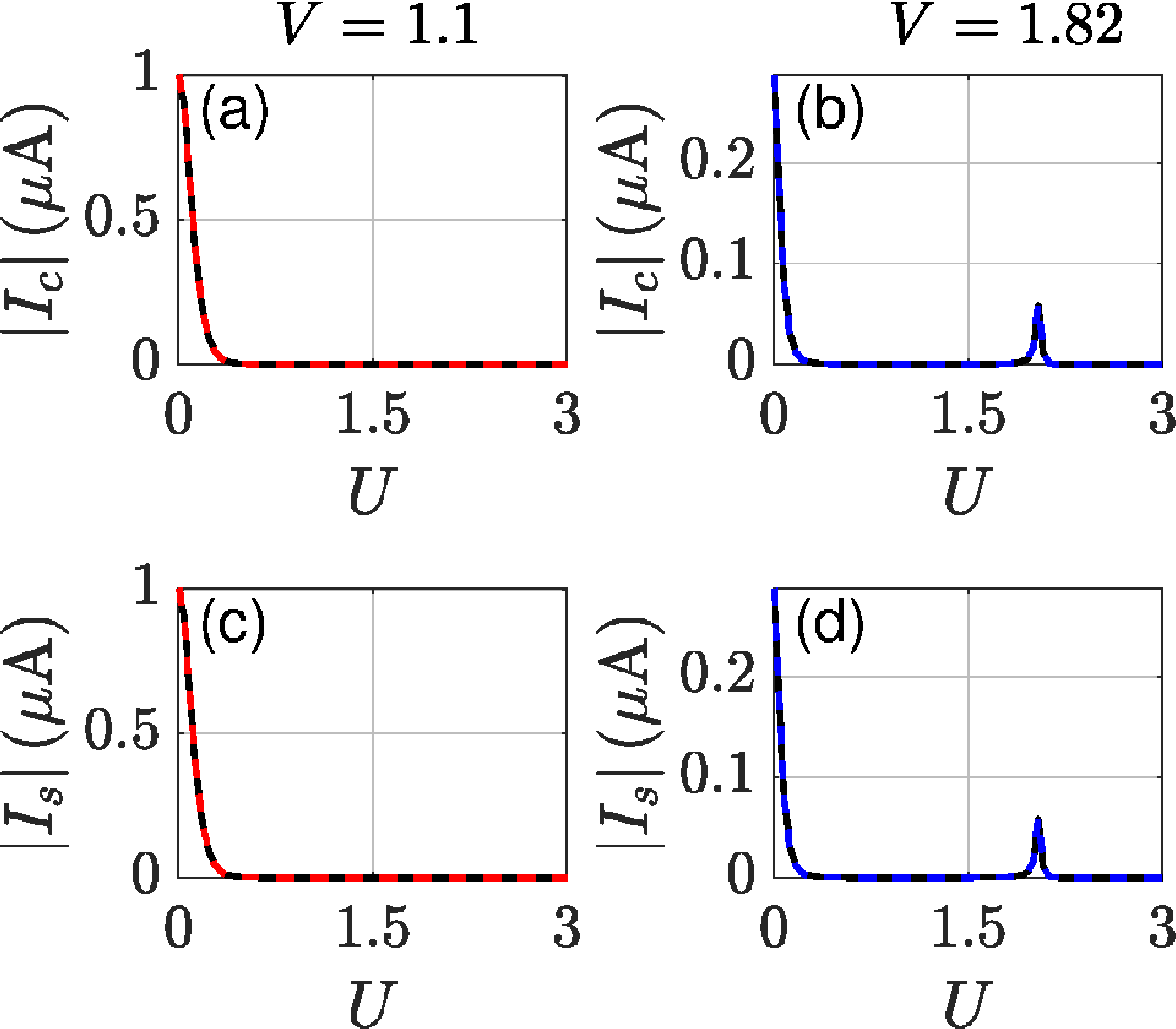}
    \caption{Interaction-strength dependence of the absolute persistent-current amplitudes at the selected modulation strengths $V$. (a) Charge current $|I_c|$ and (b) spin current $|I_s|$ as functions of the Hubbard interaction $U$ for an antiferromagnetic ring with $N=144$ sites. The hopping-modulation strength, Aharonov-Bohm flux, and staggered Zeeman field are fixed at $W=0.2$, $\phi=0.3\phi_0$, and $h_z=0.4$, respectively.}
    \label{fig:IU_DIFF_V}
\end{figure}
To clarify the contrasting interaction dependence of the primary and secondary persistent-current enhancement regimes, we examine the charge and spin currents as functions of the Hubbard interaction strength $U$ at $V=1.10$ and $V=1.82$, as shown in Fig.~\ref{fig:IU_DIFF_V}. At $V=1.10$, Figs.~\ref{fig:IU_DIFF_V}(a) and \ref{fig:IU_DIFF_V}(c) show that both $|I_c|$ and $|I_s|$ decrease sharply from their large noninteracting values as $U$ increases, becoming nearly negligible over most of the displayed range. This behavior indicates a strong suppression of the primary current peak by on-site correlations. In contrast, a qualitatively different response emerges at $V=1.82$, as illustrated in Figs.~\ref{fig:IU_DIFF_V}(b) and \ref{fig:IU_DIFF_V}(d). Although both current components initially decrease rapidly from their values at $U=0$, they exhibit a distinct secondary enhancement around $U\approx 2$, following a region in which the currents are strongly suppressed. This suppression followed by a recovery at finite interaction strength constitutes a clear re-entrant feature in the current response: the currents are initially appreciable in the noninteracting limit, become nearly quenched at intermediate $U$, and subsequently recover within a restricted interaction window. The recovered currents then decrease again as $U$ is increased further. This behavior contrasts with the predominantly monotonic suppression observed at $V=1.10$. The qualitatively similar feature in both charge and spin currents suggests that the re-entrant response is reflected in both transport channels, although the recovered current remains substantially smaller than its noninteracting value. A possible interpretation is that the Hubbard interaction modifies the electronic spectrum and the spin-resolved character of the current-carrying states, allowing their net contribution to the persistent current to recover over a limited range of $U$. The microscopic origin of this recovery requires further investigation of the interaction-dependent spectrum and eigenstate-resolved current contributions. Overall, these results distinguish the primary peak, which is rapidly suppressed by increasing $U$, from the secondary peak, which exhibits a pronounced nonmonotonic, re-entrant current response.
\begin{figure}[t]
    \centering
    \includegraphics[width=8cm,height=8cm]{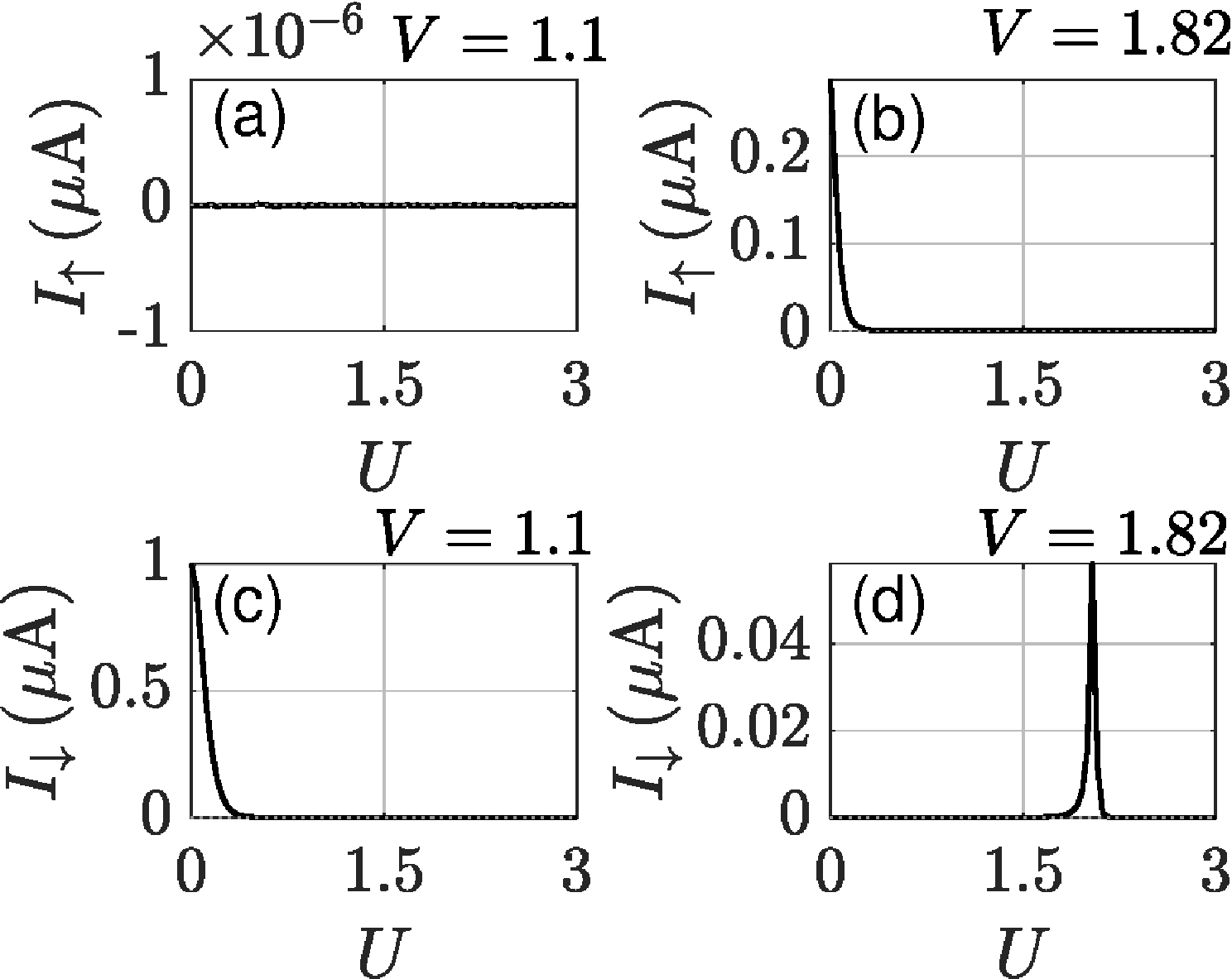}
    \caption{Spin-resolved persistent currents as functions of the Hubbard interaction strength $U$ at the two selected modulation strengths, $V=1.1$ and $V=1.82$. The spin-up current $I_{\uparrow}$ and spin-down current $I_{\downarrow}$ are shown for an antiferromagnetic ring with $N=144$ sites at half filling. The remaining parameters are fixed at $W=0.2$, $\phi=0.3\phi_0$, and $h_z=0.4$.}
    \label{fig:IT_IU}
\end{figure}

To further elucidate the re-entrant persistent-current response associated with the secondary peak, we investigate the total current $I_T$ and the spin-resolved current $I_\downarrow$ as functions of the Hubbard interaction $U$ at $V=1.10$ and $V=1.82$, as shown in Fig.~\ref{fig:IT_IU}. At $V=1.10$, the total current in Fig.~\ref{fig:IT_IU}(a) is extremely small, remaining on the order of $10^{-6}\,\mu\mathrm{A}$ throughout the displayed interaction range, whereas the spin-resolved current in Fig.~\ref{fig:IT_IU}(c) is appreciable only near $U=0$ and decreases rapidly as the interaction increases. This behavior indicates that the primary current-enhancement regime is strongly suppressed by the introduction of Hubbard correlations. In contrast, at $V=1.82$, the total current in Fig.~\ref{fig:IT_IU}(b) initially decreases sharply from its noninteracting value but exhibits a pronounced recovery at finite interaction strength, while Fig.~\ref{fig:IT_IU}(d) reveals a sharp enhancement of the spin-resolved current near $U\approx 2$. The simultaneous appearance of these finite-$U$ features provides further evidence that the secondary current-enhancement regime exhibits a re-entrant response: the current is initially finite at weak interaction, becomes strongly suppressed at intermediate $U$, and subsequently recovers within a narrow interaction window. The contrast between the two modulation strengths suggests that this recovery is specific to the electronic states associated with the secondary peak rather than a general consequence of increasing $U$. A possible interpretation is that the Hubbard interaction modifies the spin-dependent energy levels and their flux sensitivity, allowing a subset of current-carrying states to regain a substantial contribution over a restricted range of interaction strengths. The sharpness of the enhancement near $U\approx 2$ suggests that this recovery occurs within a narrow parameter window, although its precise microscopic origin requires further analysis of the spin-resolved spectrum and current contributions. Overall, these results support a spin-dependent, interaction-induced re-entrant current response associated with the secondary peak, in contrast to the strong suppression observed at the primary peak.

\subsubsection{Hubbard Interaction Effects at Quarter Filling for a Finite System}
\label{subsec:U_QUARTER}
\begin{figure}[t]
    \centering
    \includegraphics[width=8.5cm,height=4cm]{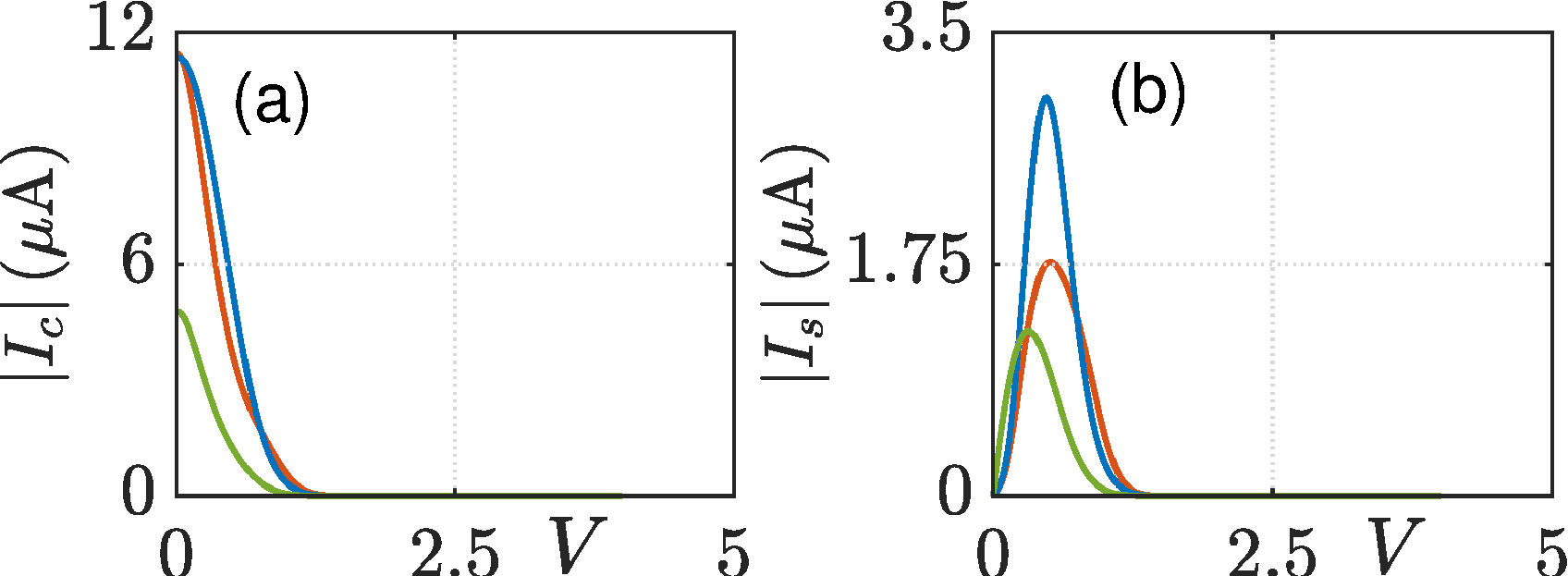}
    \caption{Absolute charge and spin persistent currents as functions of the site-potential modulation strength $V$ for three values of the Hubbard interaction, $U=0,1,$ and $2$, at quarter filling. The system size is restricted to $N=34$ owing to numerical convergence limitations at quarter filling. The remaining parameters are fixed at $W=0.2$, $\phi=0.3\phi_0$, and $h_z=0.4$.}
    \label{fig:U_QUARTER}
\end{figure}
At quarter filling, the self-consistent calculations become increasingly difficult to converge as the system size increases. We therefore restrict the present analysis to $N=34$, for which converged solutions can be obtained, and investigate the influence of the on-site Hubbard interaction $U$ on the charge and spin persistent currents as functions of the modulation strength $V$, as shown in Fig.~\ref{fig:U_QUARTER}. Both current components exhibit pronounced maxima at weak modulation and decrease rapidly with increasing $V$, indicating a progressive suppression of the persistent-current response. The charge current has a comparatively large peak amplitude, while the spin current displays a distinct dependence on the interaction strength. As $U$ increases, the peak amplitudes and their positions change, demonstrating that electron correlations modify the contributions of the current-carrying states and their sensitivity to the applied Aharonov-Bohm flux. In particular, the charge-current response is progressively suppressed, whereas the spin-current peak exhibits a nonuniform variation with $U$, reflecting the different ways in which interactions modify the spin-resolved contributions to transport. Unlike the half-filled case, where a secondary current-enhancement feature appears over a restricted range of $V$, the quarter-filled results for $N=34$ are dominated by a primary peak followed by a rapid decay, without a comparably pronounced secondary enhancement. This difference may reflect the combined influence of filling, system size, and interaction-induced changes in the electronic states. Since the quarter-filled and half-filled results correspond to different system sizes, however, their comparison does not isolate the effect of filling alone. Further calculations for larger, converged quarter-filled systems are required to establish the robustness of these trends and clarify the interplay between electron correlations and quasiperiodic modulation.

\subsubsection{Drude Weight and Re-entrant Transport Characteristics}
\label{subsec:drude}
\begin{figure}[t]
    \centering
    \includegraphics[width=8cm,height=7cm]{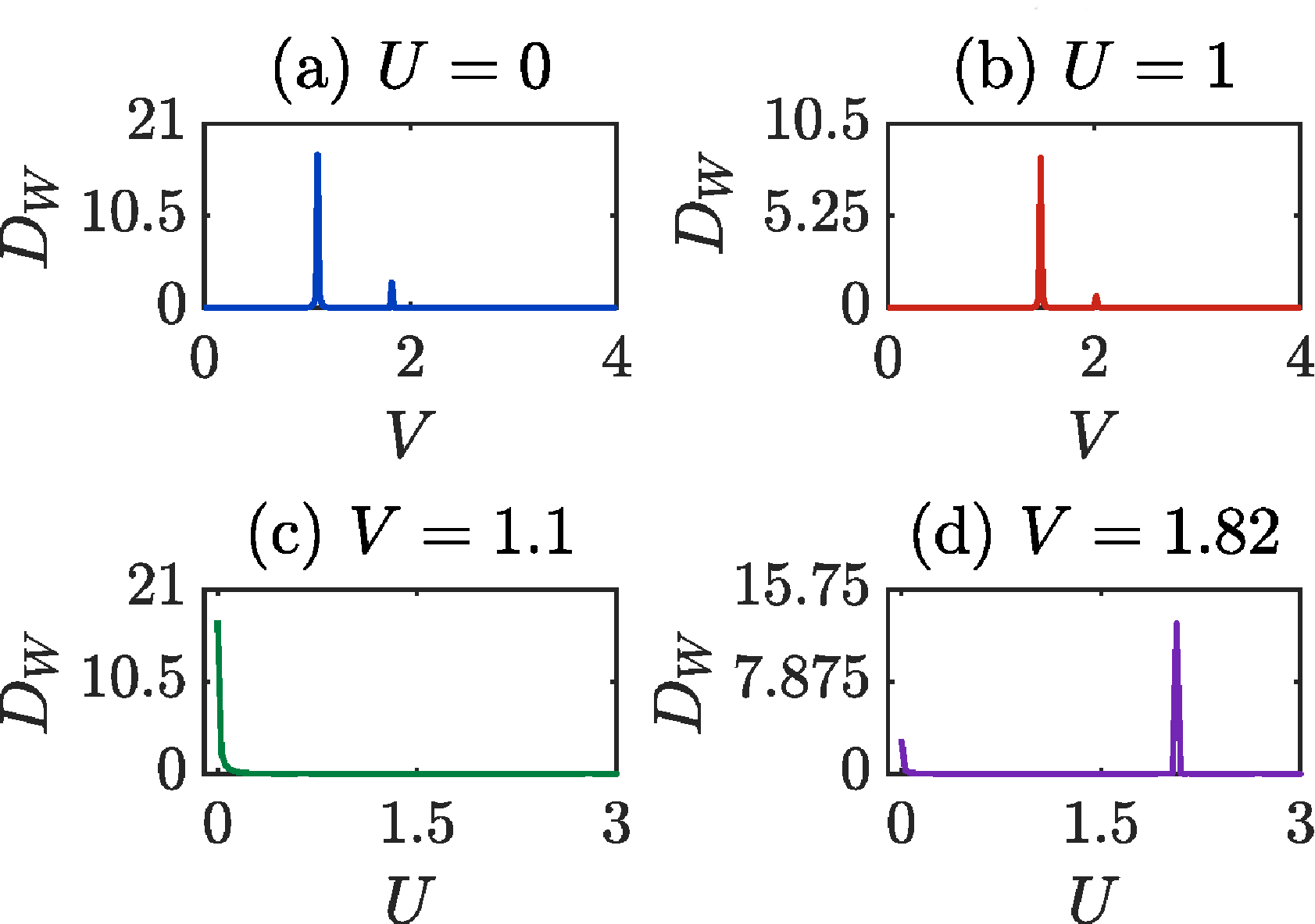}
    \caption{Variation of the Drude weight with (a)–(b) the modulation strength $V$ for two different values of the Hubbard interaction $U$, and (c)–(d) the Hubbard interaction $U$ for two different values of $V$, for a ring of $N=144$ sites with $W=0.2$.}
    \label{fig:drude}
\end{figure}

Figure~\ref{fig:drude} presents the variation of the Drude weight $D_W$ with the modulation strength $V$ for two representative interaction strengths, $U=0$ and $U=1$, in Figs.~\ref{fig:drude}(a) and \ref{fig:drude}(b), respectively. Both panels exhibit a pronounced primary peak accompanied by a weaker secondary feature, indicating a non-monotonic response of the Drude weight to the modulation strength. In particular, the secondary peak in the first-row panels is important because it reveals an additional parameter window in which the charge stiffness is enhanced, beyond the dominant peak. The peak positions and amplitudes depend on the Hubbard interaction, demonstrating that electron correlations modify the sensitivity of the electronic transport to the quasiperiodic modulation. The bottom-row panels further elucidate this interplay by showing the evolution of $D_W$ with $U$ at fixed modulation strengths $V=1.1$ and $V=1.82$. For $V=1.1$, the Drude weight is strongly enhanced near weak interaction and subsequently becomes suppressed over most of the interaction range. By contrast, at $V=1.82$, the response displays a pronounced peak at finite $U$, following a region of relatively small Drude weight. This non-monotonic recovery of the Drude weight constitutes a re-entrant feature in the interaction-dependent transport response.

The secondary peak in the first-row panels and the re-entrant enhancement in Fig.~\ref{fig:drude}(d) are consistent with the corresponding variation of the current with $U$ presented in Fig.~\ref{fig:IU_DIFF_V}. In particular, the enhancement at finite interaction for $V=1.82$ indicates that increasing $U$ does not necessarily suppress transport monotonically; instead, the interplay between the interaction and quasiperiodic modulation can restore the charge stiffness within a restricted parameter window. The correspondence between the Drude-weight features and the current response provides complementary evidence for this non-monotonic transport behavior. While the Drude weight characterizes the curvature of the ground-state energy with respect to magnetic flux and hence the charge stiffness, the current reflects the response of the system to the applied flux. Their similar parameter-dependent features support the interpretation that the observed secondary and re-entrant enhancements arise from the underlying electronic response to the combined effects of quasiperiodicity and Hubbard interaction.

\subsection{Validation of the Mean-Field Approach Against ED and DMRG}
\label{subsec:MFT_ED_DMRG}
\begin{figure}[t]
    \centering
    \includegraphics[width=8.5cm,height=4cm]{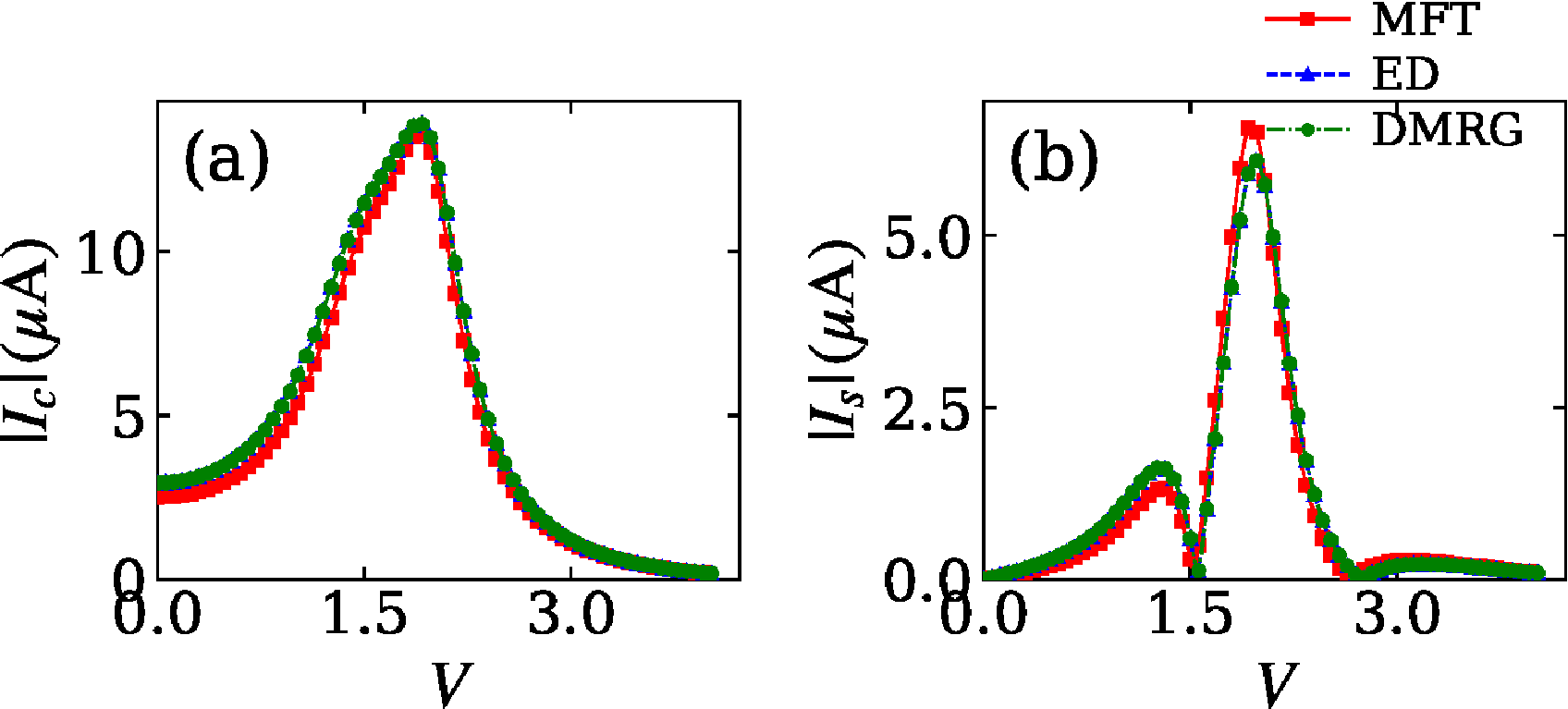}
    \caption{Comparison of the mean-field theory (MFT), exact diagonalization (ED), and density-matrix renormalization group (DMRG) results for the absolute charge and spin persistent currents as functions of the site-potential modulation strength $V$ for a 12-site antiferromagnetic ring at half filling, with Hubbard interaction strength $U=1$. The remaining parameters are fixed at $W=0.2$, $h_z=0.4$, and $\phi=0.3\phi_0$.}
    \label{fig:MFT_ED_DMRG}
\end{figure}

To establish the reliability of our mean-field theory (MFT), we compare its predictions with those obtained using exact diagonalization (ED) and density-matrix renormalization group (DMRG) calculations~\cite{r67,r68,r66} for a system of $N=16$ sites at half filling, as shown in Fig.~\ref{fig:MFT_ED_DMRG}. The comparison is performed by examining the charge and spin persistent currents as functions of the modulation strength $V$. As evident from both panels, the MFT results closely reproduce the overall trends obtained from ED and DMRG, including the pronounced current maxima and their positions along the $V$ axis. The agreement in the peak profiles and the subsequent suppression of the currents demonstrates that the mean-field description captures the principal features of the persistent-current response in the parameter regime considered. Although small quantitative differences in the peak heights and profile widths may arise from the approximate treatment of interactions within MFT and the distinct numerical approaches employed by ED and DMRG, the close correspondence among the three methods provides an important benchmark for our calculations. This validation supports the use of MFT to investigate larger systems and broader parameter ranges, where fully interacting calculations become computationally demanding. We emphasize, however, that the comparison establishes the reliability of MFT for the system size and parameters examined.

\subsection{Possible Experimental Realization}
\label{subsec:experimental_realization}

The quasiperiodic lattice considered in this work could be engineered using ultracold fermionic atoms confined in an optical lattice formed by interfering laser beams. In such a setup, a primary optical standing wave establishes the underlying one-dimensional lattice, while additional laser fields with incommensurate wavelengths or suitably controlled relative phases generate the quasiperiodic Aubry-Andr\'e-Harper (AAH) onsite potential. The staggered component can be introduced by appropriately designing the optical superlattice~\cite{exp1} whereas spatially varying tunneling amplitudes can be engineered by controlling the depths and barriers between neighboring lattice sites. Optical waveguides or spatially structured laser fields can provide additional control over the geometry and effective coupling between sites~\cite{exp2}. Two internal atomic states represent the spin degrees of freedom, and spin-dependent optical potentials or Raman-assisted~\cite{exp5} couplings can produce an effective staggered Zeeman field corresponding to the antiferromagnetic configuration in the model. The onsite Hubbard interaction can be tuned through atom-atom interactions, with Feshbach-resonance techniques providing additional control where experimentally available. A ring geometry can be realized by confining the atoms in a toroidal optical potential, while rotation or laser-induced synthetic gauge fields introduce an effective Aharonov-Bohm flux. This arrangement offers a controllable platform for exploring the effects of quasiperiodic onsite and hopping modulations, staggered magnetic fields, and interactions on the persistent-current response.

For a complementary implementation in a mesoscopic electronic ring, the quasiperiodic onsite landscape and spatially modulated hopping amplitudes could be approximated using patterned nanostructures, electrostatic gates, and tunable tunnel barriers. Magnetic proximity coupling or engineered magnetic interfaces could provide an effective staggered spin-dependent field, while electron-electron interactions arise from Coulomb repulsion in the confined geometry. An external magnetic field threading the ring introduces the AB phase, and the associated persistent-current response can be probed through its magnetic signature\cite{exp3} using a superconducting quantum interference device (SQUID)~\cite{exp4} coupled to the ring. By varying the magnetic flux and gate-controlled parameters, one could investigate the predicted current-enhancement regimes and their evolution with interaction strength. Accessing the spin-resolved currents would additionally require a spin-sensitive detection method. These two approaches offer complementary possibilities: optical lattices provide flexibility in engineering the quasiperiodic Hamiltonian and tuning interactions, whereas mesoscopic rings offer a direct route to detecting flux-dependent persistent currents through sensitive magnetometry.
\begin{figure*}[t]
    \centering
    \includegraphics[width=18cm,height=10cm]{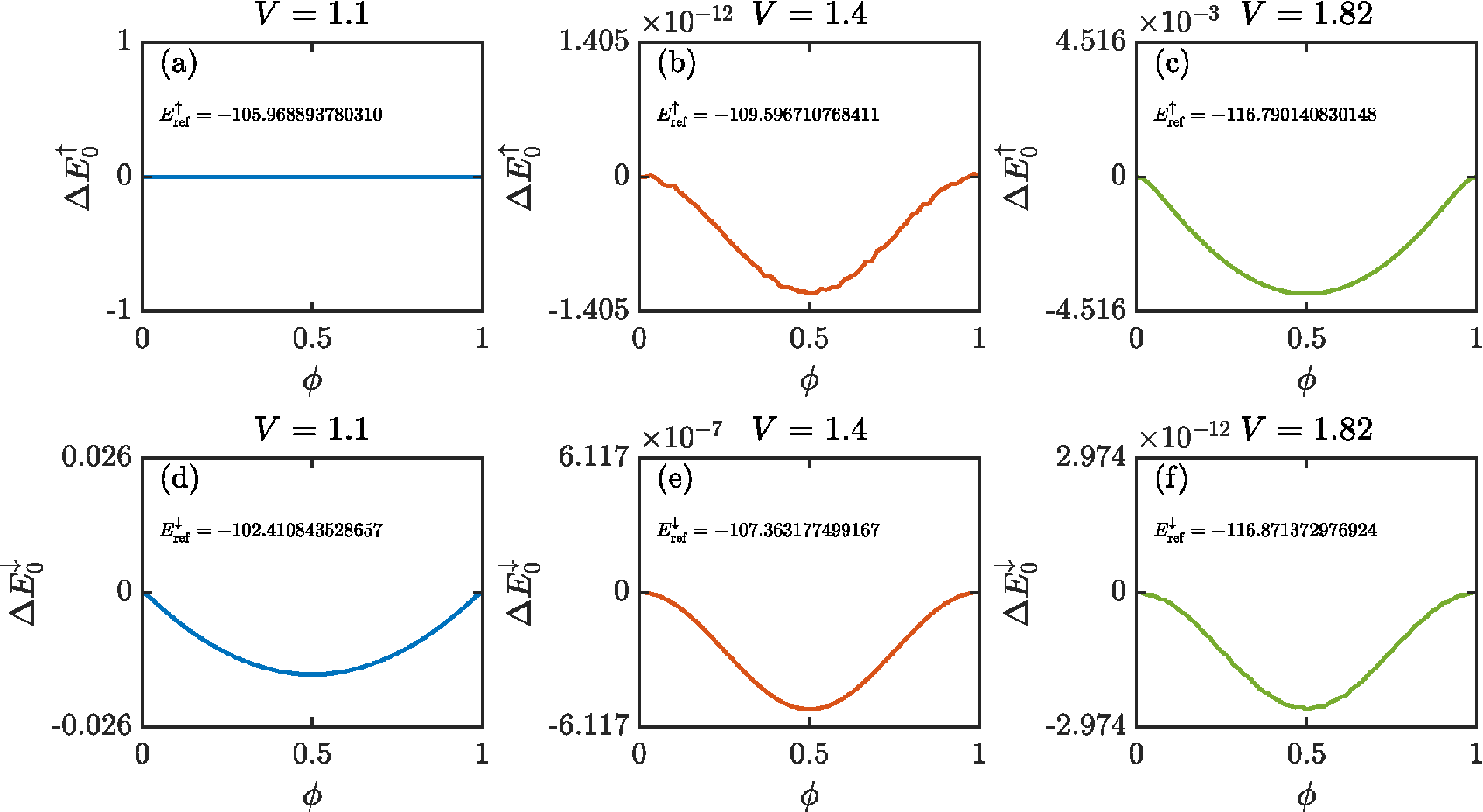}
    \caption{Ground-state energies of the spin-up (upper row) and spin-down (lower row) sectors as functions of the Aharonov-Bohm flux $\phi$ for three representative values of the site-potential modulation strength $V$, corresponding to the primary peak, secondary peak, and an additional reference value. Results are shown for an antiferromagnetic ring with $N=144$ sites at $U=0$, $W=0.2$, and $h_z=0.4$.}
    \label{fig:GS}
\end{figure*}

\section{Conclusion}

In this work, we have investigated how generalized Aubry-Andr\'e-Harper (AAH) modulation with staggered onsite energies, together with antiferromagnetic (AFM) ordering, influences persistent charge and spin currents in a one-dimensional quantum ring, both in the absence and in the presence of on-site Hubbard interactions. At half filling, two distinct current-enhancement peaks emerge near $V=1.1$ and $V=1.82$, whereas at quarter filling, the spin-current response exhibits only a single peak, underscoring the influence of particle filling on the current characteristics. The combined effects of staggered AAH modulation and AFM ordering modify the spin-resolved energy spectrum and redistribute the magnetic-flux sensitivity between the two spin sectors. We further establish that quasiperiodic hopping modulation is essential for the emergence of the two-peak structure at half filling. The inverse participation ratio (IPR), spin-resolved bond-current profiles, and flux-dependent ground-state energies provide complementary insights into the eigenstate properties, the dominant spin contributions, and the flux response of the system.

The inclusion of Hubbard interactions leads to distinct changes in the two current-enhancement regimes. While the primary peak is suppressed, the secondary peak exhibits a nonmonotonic, re-entrant response: the current decreases with increasing $U$, becomes strongly suppressed at intermediate interaction strengths, and subsequently recovers around $U\approx 2$. The Drude weight further characterizes this interaction-dependent behavior. Importantly, the Hartree-Fock mean-field theory (MFT) results agree well with exact diagonalization (ED) and density-matrix renormalization group (DMRG) calculations for a finite half-filled system, lending support to the reliability of the mean-field description in the parameter regime examined. Overall, our findings demonstrate that the interplay of quasiperiodic onsite and hopping modulations, AFM ordering, particle filling, and electronic correlations provides a means of controlling the magnitude and spin dependence of persistent currents, including the emergence of multiple enhancement peaks and interaction-induced re-entrant behavior.

\appendix
\section{Flux Dependence of the Spin-Resolved Ground-State Energy}
 \label{subsec:GS_energy_flux}
 In this section, we examine the flux dependence of the spin-resolved ground-state energies to elucidate the origin of the modulation-dependent persistent-current response at half filling. We consider the spin-up and spin-down sectors for $V=1.1$, $1.4$, and $1.8$, as shown in Fig.~\ref{fig:GS}. The upper and lower rows correspond to the spin-up and spin-down sectors, respectively, while the three columns represent the different values of the quasiperiodic modulation strength. To resolve the small flux-induced variations more clearly, we plot the relative ground-state energies, $\Delta E_0^{\sigma}(\phi)=E_0^{\sigma}(\phi)-E_{\mathrm{ref}}^{\sigma}$, where $\sigma=\uparrow,\downarrow$ and $E_{\mathrm{ref}}^{\sigma}=E_0^{\sigma}(\phi=0)$ denotes the reference energy in each spin sector for the corresponding value of $V$. Consequently, $\Delta E_0^{\sigma}(0)=0$ by construction, and the plotted curves directly display the energy variation relative to the zero-flux configuration. The vertical-axis ranges are adjusted individually to make these small variations visible, particularly in panels (b)–(f); this rescaling affects only the visual presentation and does not alter the calculated energies or their flux dependence.

The persistent current associated with each spin sector is determined by the flux derivative of its ground-state energy, $I_{\sigma}(\phi)=-\partial E_0^{\sigma}/\partial\phi$, up to the convention-dependent prefactor associated with the definition of the dimensionless flux. Since the reference energy is independent of $\phi$, subtracting $E_{\mathrm{ref}}^{\sigma}$ leaves this derivative unchanged. Thus, the amplitude and local slope of $\Delta E_0^{\sigma}(\phi)$ provide a direct indication of the relative flux sensitivity of the corresponding spin sector. For the spin-up sector, the flux-dependent variation is most pronounced at $V=1.8$, whereas the variations at $V=1.1$ and $1.4$ are substantially smaller and become apparent only upon resolving the energy differences at approximately the eighth or ninth decimal place. In contrast, the spin-down sector exhibits its strongest flux dependence at $V=1.1$, followed by much weaker variations at $V=1.4$ and $1.8$. The reference energies $E_{\mathrm{ref}}^{\uparrow}$ and $E_{\mathrm{ref}}^{\downarrow}$, indicated within the corresponding panels, specify the zero-flux energy baselines from which these differences are measured; they should not be confused with the plotted relative energies.

These results reveal a contrasting modulation dependence of the two spin sectors: increasing $V$ from $1.1$ to $1.8$ enhances the flux sensitivity of the spin-up sector while suppressing that of the spin-down sector. The dominant spin-sector contribution to the persistent-current response therefore shifts from spin-down at $V=1.1$ to spin-up at $V=1.8$. The relative-energy representation makes this redistribution particularly clear by separating the small flux-induced variations from the much larger absolute ground-state-energy offsets. Overall, the spin-resolved energy profiles provide complementary evidence that the quasiperiodic modulation controls the flux sensitivity of the two spin sectors in different ways, thereby shaping the modulation-dependent persistent-current response.

\begin{acknowledgments}
We thank Dr. Soumya Ranjan Padhi for useful discussions. We gratefully acknowledge financial support from the Institute Postdoctoral Fellowship at IIT Bhubaneswar and the postdoctoral fellowship at the Institute of Physics (IOP), Bhubaneswar.
\end{acknowledgments}

\bibliography{reference}

\end{document}